\documentclass[12pt, nofootinbib]{article}

\usepackage{amssymb}
\usepackage{amsmath,bm}
\usepackage{amssymb}
\usepackage{graphicx}
\usepackage{amsfonts}         
\usepackage{fancybox}

\usepackage[usenames,dvipsnames]{xcolor}

\usepackage[pagebackref,  
bookmarks={false}, pdfauthor={John McGreevy}, pdftitle={Yay, physics!}]{hyperref}
\hypersetup{colorlinks=true, 
linkcolor=BrickRed, 
citecolor=Violet, 
filecolor=OliveGreen, 
urlcolor=RoyalBlue, 
filebordercolor={.8 .8 1}, 
urlbordercolor={.8 .8 0}}
\usepackage{soul}
\setstcolor{Red}

\definecolor{darkgreen}{rgb}{0,0.4,0}
\definecolor{darkred}{rgb}{0.4,0,0}
\definecolor{darkblue}{rgb}{0,0,0.4}
\definecolor{lightblue}{rgb}{.6,.6,0.9}

\definecolor{uglybrown}{rgb}{0.8,  0.7,  0.5}

\definecolor{palatinatepurple}{rgb}{0.41, 0.16, 0.38}
\definecolor{celebrationcolor}{rgb}{0.75,  0.0,  0.9}

\definecolor{shadecolor}{rgb}{0.90,0.90,0.90}
\definecolor{DVcolor}{rgb}{0.95,  0.5,  0.2}
\definecolor{lightbluemuons}{rgb}{0.0,.65,1.0}

\definecolor{chartreuse}{rgb}{0.70, 1.00, 0.00}

\usepackage{tikz}
\usetikzlibrary{shapes}
\usetikzlibrary{trees}
\usetikzlibrary{snakes}
\usetikzlibrary{matrix,arrows} 				
\usetikzlibrary{positioning}				
\usetikzlibrary{calc,through}				
\usetikzlibrary{decorations.pathreplacing}  
\usepackage[tikz]{bclogo} 					
\usepackage{pgffor}							

\usetikzlibrary{decorations.markings}
\usetikzlibrary{intersections}				

\usetikzlibrary{arrows,decorations.pathmorphing,backgrounds,positioning,fit,petri,automata,shadows,calendar,mindmap, graphs}
\usetikzlibrary{arrows.meta,bending}

\tikzset{
    vector/.style={decorate, decoration={snake}, draw},
    fermion/.style={postaction={decorate},
        decoration={markings,mark=at position .55 with {\arrow{>}}}},
    fermionbar/.style={draw, postaction={decorate},
        decoration={markings,mark=at position .55 with {\arrow{<}}}},
    fermionnoarrow/.style={},
    gluon/.style={decorate,
        decoration={coil,amplitude=4pt, segment length=5pt}},
    scalar/.style={dashed, postaction={decorate},
        decoration={markings,mark=at position .55 with {\arrow{>}}}},
    scalarbar/.style={dashed, postaction={decorate},
        decoration={markings,mark=at position .55 with {\arrow{<}}}},
    scalarnoarrow/.style={dashed,draw},
	vectorscalar/.style={loosely dotted,draw=black, postaction={decorate}},
}

\def\centerarc[#1](#2)(#3:#4:#5)
    { \draw[#1] ($(#2)+({#5*cos(#3)},{#5*sin(#3)})$) arc (#3:#4:#5); }

\usepackage{enumitem}

\usepackage{slashed}

\usepackage[font=small,labelfont=bf]{caption}
\DeclareCaptionFont{tiny}{\tiny}
\usepackage{epstopdf}
\usepackage{wasysym}

\usepackage[yyyymmdd,hhmmss]{datetime}   
\usepackage{framed}  
 \usepackage{mdframed}
 \usepackage{wrapfig}
 \usepackage{yfonts}  

\def\ketbra#1#2{ | #1 \rangle\hskip-2pt\langle #2|}

\newmdenv[%
        backgroundcolor=lightgray,
    linecolor=black,
    outerlinewidth=2pt,
]{boxedandshaded}

\def\thump{\hfill$\blacksquare$}

\def\parfig#1#2{
\parbox{#1\textwidth}
{\includegraphics[width=#1\textwidth]{#2}}
}

\numberwithin{equation}{section}

\def\nd{{ \vphantom{\dagger}}}

\newcommand{\vev}[1]{\langle #1 \rangle}

\newlength{\extraspace}
\newlength{\extraspaces}
\def\be{\begin{equation}}
\def\ee{\end{equation}}

\newcommand{\bea}{\begin{eqnarray}}
\newcommand{\eea}{\end{eqnarray}}

\def\half{{1\over 2}}

\def\tr{{\rm tr}}

\def\bra#1{\left\langle#1\right|}
\def\ket#1{\left|#1\right\rangle}

\def\vev#1{\left\langle{#1}\right\rangle}

\def\CE{{\cal E}}

\def\CH{{\cal H}}
\def\CI{{\cal I}}

\def\CO{{\cal O}}

\def\II{\relax{I\kern-.10em I}}

\def\IB{\relax{\rm I\kern-.18em B}}

\def\ID{\relax{\rm I\kern-.18em D}}
\def\IE{\relax{\rm I\kern-.18em E}}
\def\IF{\relax{\rm I\kern-.18em F}}
\def\IG{\relax\hbox{$\inbar\kern-.3em{\rm G}$}}
\def\IGa{\relax\hbox{${\rm I}\kern-.18em\Gamma$}}
\def\IH{\relax{\rm I\kern-.18em H}}
\def\II{\relax{\rm I\kern-.18em I}}
\def\IK{\relax{\rm I\kern-.18em K}}

\def\inbar{\,\vrule height1.5ex width.4pt depth0pt}

\def\lp10{\ell_p^{10}}
\def\lp11{\ell_p^{11}}
\def\R11{R_{11}}

\def\frac#1#2{{#1 \over #2}}

\def\up{\uparrow}
\def\down{\downarrow}

\def\Ione{\hbox{$1\hskip -1.2pt\vrule depth 0pt height 1.53ex width 0.7pt
                  \vrule depth 0pt height 0.3pt width 0.12em$}}

\newdimen\tableauside\tableauside=1.0ex
\newdimen\tableaurule\tableaurule=0.4pt
\newdimen\tableaustep
\def\phantomhrule#1{\hbox{\vbox to0pt{\hrule height\tableaurule width#1\vss}}}
\def\phantomvrule#1{\vbox{\hbox to0pt{\vrule width\tableaurule height#1\hss}}}
\def\sqr{\vbox{%
  \phantomhrule\tableaustep
  \hbox{\phantomvrule\tableaustep\kern\tableaustep\phantomvrule\tableaustep}%
  \hbox{\vbox{\phantomhrule\tableauside}\kern-\tableaurule}}}
\def\squares#1{\hbox{\count0=#1\noindent\loop\sqr
  \advance\count0 by-1 \ifnum\count0>0\repeat}}
\def\tableau#1{\vcenter{\offinterlineskip
  \tableaustep=\tableauside\advance\tableaustep by-\tableaurule
  \kern\normallineskip\hbox
    {\kern\normallineskip\vbox
      {\gettableau#1 0 }%
     \kern\normallineskip\kern\tableaurule}%
  \kern\normallineskip\kern\tableaurule}}
\def\gettableau#1 {\ifnum#1=0\let\next=\null\else
  \squares{#1}\let\next=\gettableau\fi\next}

\def\({\left(}
\def\){\right)}

\def\ii{{\bf i}}

\def\HH{{\bf H}}

\def\lsim{\mathrel{\mathstrut\smash{\ooalign{\raise2.5pt\hbox{$<$}\cr\lower2.5pt\hbox{$\sim$}}}}}
\def\gsim{\mathrel{\mathstrut\smash{\ooalign{\raise2.5pt\hbox{$>$}\cr\lower2.5pt\hbox{$\sim$}}}}}

\def\overleftrightarrow#1{\vbox{\ialign{##\crcr
     $\leftrightarrow$\crcr\noalign{\kern-0pt\nointerlineskip}
     $\hfil\displaystyle{#1}\hfil$\crcr}}}
     
     \def\overleftarrow#1{\vbox{\ialign{##\crcr
     $\leftarrow$\crcr\noalign{\kern-0pt\nointerlineskip}
     $\hfil\displaystyle{#1}\hfil$\crcr}}}

\def\ie{{\it i.e.}}

\newif{\ifeq}           
\eqtrue                 
\newcounter{lecturecounter}

\def\baselinestretch{1.1}
\renewcommand{\title}[1]{\vbox{\center\LARGE{#1}}\vspace{5mm}}
\renewcommand{\author}[1]{\vbox{\center#1}\vspace{5mm}}
\newcommand{\address}[1]{\vbox{\center\em#1}}

\renewcommand{\date}[1]{\vbox{\center#1}}

\usepackage{cite}

\begin{document}

\title{
Disentangling quantum matter with measurements
}

\author{
Daniel Ben-Zion, John McGreevy, Tarun Grover}

\address{Department of Physics, University of California at San Diego, La Jolla, CA 92093, USA}

\begin{abstract}
Measurements destroy entanglement.  
Building on ideas used to study `quantum disentangled liquids', 
we explore the use of this effect to
characterize states of matter.  We focus on systems with multiple components, such as charge and spin
in a Hubbard model, or local moments and conduction electrons in a Kondo lattice model.
In such systems, measurements of (a subset of) one of the components can leave behind a quantum state of 
the other that is easy to understand, for example in terms of scaling of entanglement entropy of subregions.  
We bound the outcome of this protocol, for any choice of measurement, in terms of more standard information-theoretic quantities.  
We apply this quantum disentangling protocol to several problems of physical interest, including gapless topological phases, heavy fermions, and scar states in Hubbard model.

\end{abstract}

December 2019, revised February 2020

\vfill\eject

\renewcommand{\baselinestretch}{0.75}\normalsize
\tableofcontents
\renewcommand{\baselinestretch}{1.1}\normalsize

\vfill\eject

\section{Introduction}
\label{sec:intro}

Consider a quantum state of a two-component system, say, local moments interacting with electrons as in a Kondo lattice model, or spin and charge degrees of freedom in a Hubbard model. How shall one characterize the quantum entanglement between the two components? One possible route is to integrate out one of the components, and study the properties of the resulting reduced density matrix corresponding to the other component. Now one can characterize the entanglement between the two components via the von Neumann entropy of the resulting reduced density matrix. For example, if the two components were unentangled to begin with, then the reduced density matrix would be pure, and therefore will have zero von Neumann entropy. At the other extreme, if the original wavefunction 
satisfies the eigenstate thermalization hypothesis (ETH), then the resulting density matrix will have a volume law von Neumann entropy \cite{deutsch1991, srednicki1994chaos, rigol2008, rigol2012, Deutsch2010, rigol_review, garrison2015does}. A seemingly very different, and perhaps more feasible approach from an experimental standpoint, is to perform a \textit{measurement} on only one of the components such that the state of that component is fully specified, and study the resulting wavefunction. Basic principles of quantum mechanics dictate that the measurement renders the measured degrees of freedom as classical objects with definite values, and now the only quantum degrees of freedom belong to the unmeasured component. One measure of the entanglement between the two components  is the \textit{change} in the entanglement between different subsystems of the unmeasured component due to the measurement. For example, if the two components were unentangled to begin with, then the measurement leaves the reduced density matrix of the unmeasured component completely unchanged. 

One application of such a partial-measurement-based protocol was discussed in Ref.~\cite{grover2014quantum}, where a new state of matter, called `quantum disentangled liquid' (QDL), was introduced. In a QDL state, measurement leads to a dramatic reduction in the  bipartite entanglement of the unmeasured component (`disentangling'). Specifically, in such a phase, although the original wavefunction has a volume law bipartite entanglement, the unmeasured component only has area-law entanglement post measurement. It was argued that in contrast, in a conventional `non-QDL' system, a similar protocol will instead lead to volume law entanglement for the unmeasured component post-measurement. 
A physical picture \cite{grover2014quantum} is that one component consists of `heavy' particles, whose positions provide a disorder potential which can Anderson/many-body localize the `light' particles.
A context in which the physics of QDL is realized is the Hubbard model in $1+1$ dimensions
\cite{veness2017quantum, veness2017atypical}, where the role of `heavy' and `light' particles is played by 
the spin and charge degrees of freedom respectively.
Strong numerical evidence was found that
a band of QDL-like states survives the breaking of integrability \cite{garrison2017partial}.
These are examples of `scar states' -- states in the middle of the spectrum of a non-integrable system that are not ergodic
(in a many-body system, this means that they violate ETH).

In this paper we will show that the two seemingly-different ways to characterizing entanglement introduced above - integrating out versus partial measurement - are intimately related. 
We will bound the outcome of the QDL protocol in terms of various {\it conditional information measures} -- 
combinations of von Neumann entropies of subsystems, which can be interpreted as a quantum analog of conditioning 
on the subset of measured degrees of freedom.
Specifically, we will show in \S\ref{subsec:bound} that a specific kind of conditional entropy  
provides a lower bound on the expected entanglement of a state after a partial measurement. 
We also give (less-effective) upper and lower bounds on an alternate version of the QDL protocol in terms of conditional mutual information (CMI).

One practical advantage of the conditional information measures is that, in contrast to the measurement-based protocol, they are operator agnostic -- 
they do not depend on a choice of which operator to measure.  Relatedly, in an exact diagonalization study,  implementation of conditional information measures  do not involve averaging over any degrees of freedom, in contrast to measurement-based protocol where one needs to average over the outcome of a measurement, which can be time-consuming.

As just discussed, entanglement after partial measurement has so far been 
used
as a tool to characterize scar states, or more generally, to address questions related to quantum thermalization in closed many-body systems. We will show that this set of ideas has much broader applications, and is especially useful in characterizing entanglement in ground states of multi-component systems. As an example, in 
\S\ref{sec:hcb} we will characterize a `gapless topological phase' in a model of spinful bosons where the charge degrees of freedom form a Luttinger liquid, while the spin degrees of freedom represent a symmetry protected topological (SPT) phase which is decoupled from the charge degrees of freedom at low energies. Specifically, we will show that measuring the charge degrees of freedom results in a gapped SPT wavefunction of spins, where the edge states in the entanglement spectrum are now much more apparent compared to those in the entanglement spectrum of the full gapless wavefunction. The post-measurement wavefunction can also be used to understand phase transitions in the spin-sector between a SPT phase and a non-SPT phase.

As another application to ground state wavefunctions of correlated electrons, we will characterize phases relevant to Kondo lattice systems using conditional mutual information. In \S\ref{sec:heavy-fermions}, we will show that in a heavy Fermi liquid, within a mean-field description, the entanglement of local moments conditioned on conduction electrons violates the area-law entanglement scaling in the manner of a Fermi liquid, thus exposing the underlying large Fermi surface which includes the local moments. We will also discuss the utility of conditional mutual information in diagnosing topological order in a fractionalized Fermi liquid, where at low energies local moments decouple from the conduction electrons, and become topologically ordered.

We return in \S\ref{sec:hubbard} to the notion of QDL in a Hubbard model discussed in Ref.~\cite{garrison2017partial}, and study it from the perspective of conditional information measures. We will show that the `scar states' where spin degrees of freedom effectively decouple from the charge degrees of freedom have a distinctive footprint in both conditional entropy and CMI, similar to the signature in the measurement-based diagnostic.

In \S\ref{sec:negativity} we observe that the entanglement negativity can also be used to distinguish QDL behavior from ergodicity.  
This is a measure of the bipartite entanglement in a mixed state which vanishes in separable states 
$\sum_c \rho_A^c \otimes \rho_B^c$ -- it is a measure of quantum entanglement, and not classical correlations.
A QDL state is precisely one where the entanglement of the light degrees of freedom alone is area law, whereas a general ergodic state has longer-range entanglement.  We identify a precise situation where a sharp distinction can be made,
and verify the expected behavior in the Hubbard model.

We note that the quantity called {\it SPT entanglement}, defined in \cite{2013arXiv1307.6617M} and shown there to label SPT states 
(at least for abelian groups), is an example of a measurement-based protocol similar to the ones we study.

\def\QDL{S_\text{QDL}}
\def\IQDL{I_\text{QDL}}

\subsection{A bound on the QDL diagnostic}

\label{subsec:bound}

Following the discussion in Ref.~\cite{grover2014quantum}, consider a measurement on the degrees of freedom belonging to only one component in a two-component system. A measurement effectively freezes the measured degrees of freedom to the outcome of the measurement. In the resulting wavefunction, the only quantum fluctuations correspond to the unmeasured component, and the  probabilities associated with these quantum fluctuations can therefore be thought of as \textit{conditional} probabilities - they are conditioned on the outcome of the measurement. This motivates us to seek a connection between the measurement based QDL diagnostic of Ref.~\cite{grover2014quantum}, and \textit{conditional entropy} (CE).

To build such a connection, we recall some general aspects of the QDL diagnostic. 
Consider a Hilbert space $ \mathcal H = A\otimes B \otimes C$ with three parts.
The QDL protocol 
takes as input a state $\rho_{ABC}$ and a choice of operator $X_C$ on $C$.
We will assume that the outcomes of $X_C$ provide a non-degenerate basis for $C$,
so a measurement of $X_C$ with outcome $c$ completely specifies the the state of $C$ to be $\ket{c}$.
The protocol is: 
\begin{enumerate}
\item  Measure $X_c$ and obtain outcome $c$ with 
probability $p_c = \tr_{AB} \bra{c} \rho_{ABC} \ket{c}$. 
\item In the resulting state $\rho_{AB}^c$ find the von Neumann entropy of 
subsystem $A$, $S(\rho_A^c  \equiv \tr_B \rho_{AB}^c)$\footnote{Since $\rho_{AB}^c$ is a mixed state,
this includes entropy of mixture, in addition to entropy of entanglement.  
With this in mind, we study instead a measure of mixed-state entanglement 
(the logarithmic negativity) in \S\ref{sec:negativity}.
}. 
\item  Average over the distribution $p_c$ to obtain the QDL diagnostic
$$ 
\QDL(A|X_C)
\equiv \sum_{c} p_c S(\rho^c_A).$$
\end{enumerate}

A related quantity which depends on a state $\rho_{ABC}$ but not a choice of operator 
is the {\it conditional entropy} of $A$ conditioned on $C$\footnote{For a subsystem $A$ and a density matrix $\rho$ on a larger system we denote the von Neumann entropy 
$S_\rho(A) = S(\rho_A) = - \tr \rho_A \log \rho_A$, with $\rho_A \equiv \tr_{\bar A} \rho$.  
When there is no ambiguity about the density matrix in question, we will write $S(A)$.
}
$$ S(A|C) \equiv S_\rho(AC) - S_{\rho}(C).$$
Classically, the conditional entropy is the Shannon entropy of the conditional probability distribution $p(a|c)$; 
quantum mechanically it is not the von Neumann entropy of any state, and indeed can be negative.  Its being negative 
is a sign that subsystems $A$ and $C$ are entangled.  The conditional entropy has an operational meaning 
\cite{HOW} in terms of the number of qbits from $A$ needed for $C$ to reconstruct their joint state $\rho_{AC}$ given 
free local operations and classical communication; when it is negative it means that 
their entanglement can be used as a resource, for example to teleport quantum information from $C$ to $A$.

We now show that the conditional entropy
is a lower bound for the  QDL diagnostic for any $X_C$:
\be 
S(A|C)\leq\QDL(A|X_C)
~.\label{eq:ce-bound}
\ee

Proof: 
First, we note that 
the conditional entropy can be rewritten in terms of a
{\it relative entropy}, $D(\rho|| \sigma) \equiv \tr\rho \log \rho - \tr \rho \log \sigma$,
as follows:
\be\label{eq:condition-entropy-in-terms-of-D} S(A|C) = A - D(\rho_{AC} || u_A \otimes \rho_C )  \ee
where $u_A$ is the uniform density matrix on $\CH_A$, $u_A \equiv \Ione_A/\dim \CH_A $, 
and we find it convenient to use $A \equiv \log \dim \CH_A$
to denote the size of region $A$.

The rewriting \eqref{eq:condition-entropy-in-terms-of-D}  is useful because
the relative entropy is monotonic under the action of any quantum channel $\mathcal E$
\be\label{ineq:MRE} 
~~~~D( \rho|| \sigma)  \geq D( \CE(\rho) || \CE(\sigma)) .\ee
Consider in particular the diagonal-part channel on $C$ in the basis of eigenstates $\ket{c}_C$ of $\mathcal O_C$, defined as:
\be\label{eq:diagonal-image} \rho_{ABC} \mapsto \mathcal E(\rho_{ABC} ) \equiv \sum_c \bra{c} \rho_{ABC} \ket{c} \otimes \ketbra{c}{c}
\equiv \sum_c p_c \rho_{AB}^c\otimes \ketbra{c}{c}
. \ee
This is the state that obtains if $X_C$ is measured, but the outcome of the measurement is not known.

The key step is 
\begin{align}
S(A|C)
& \buildrel{\eqref{ineq:MRE}}\over{\leq} A - D(\CE_C(\rho_{AC}) || \CE_C(u_A \otimes \rho_C) ) 
\\ & = \sum_c p_c S(\rho_A^c)  = \QDL(A|X_C)~.
\end{align}
The right equality is shown in great detail in Appendix \ref{app:bound-details}. 
Therefore,
$$ S(A|C)\leq \QDL(A|X_C)
,$$
a lower bound on the QDL quantity.
\thump

\begin{figure}[h!]
$$ \parfig{.5}{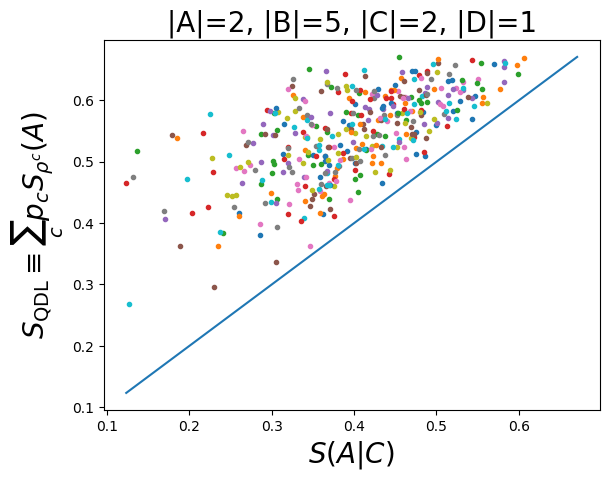}
~~\parfig{.5}{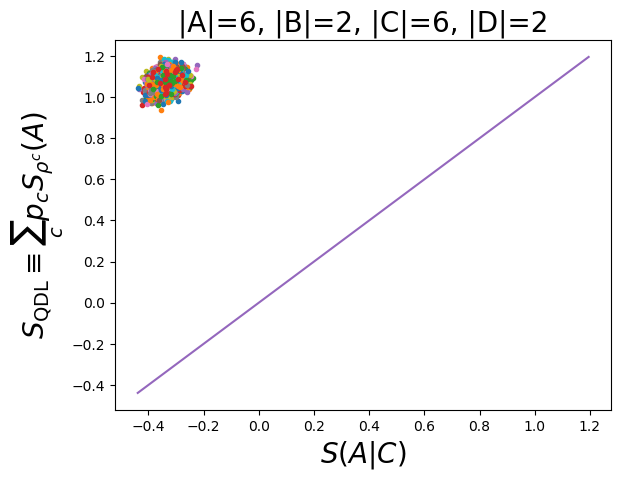}
$$
\caption{Comparison 
of the conditional entropy $S(A|C)$ with the QDL quantity 
$\QDL \equiv \sum_c p_c S_{\rho^c}(A) $ for several Haar random states of $ABCD$,
with $ABCD$ of the given dimensions.
Left: For some values of subsystem sizes, the bound is seen to be fairly tight.  
Right: In other cases, as when $S(A|C)$ is negative, the bound is loose.  Negative $S(A|C)$ is an indication that subsystems $A$ and $C$ are entangled more with each other than with the rest of the world.
\label{fig:numerical-CE-bound}
}
\label{fig:IQDL-CMI}
\end{figure}

How tight is the bound we just proved?  To learn something about this, in Fig.~\ref{fig:numerical-CE-bound} we show the QDL diagnostic versus conditional entropy
for a collection of Haar-random states, for various choices of partitions of some small Hilbert spaces.
By Haar-random states we mean states of the form $U \ket{\psi_0}$ where 
$\ket{\psi_0}$ is some reference state (here, a product state) and $U$ 
is a unitary sampled from the Haar measure.

As a side remark, we note the relevance of the notion of ``quantum discord'' \cite{2001JPhA...34.6899H, ollivier2001quantum} to this discussion.   This quantity was introduced in attempts to distinguish 
between quantum and classical correlations in a given quantum state.
Given a bipartite density matrix $\rho_{AC}$ and a measurement $X$ on $C$, 
the discord is defined to be 
\be \text{discord}(\rho_{AC}, X) \equiv S_A + S_C - S_{AC} - \chi(p_c, \rho^c) = - S(A|C) + \QDL(A|X_C), \ee
the difference between the QDL quantity 
and our lower bound for it.  In the middle step, $ \chi(p_c, \rho^c) \equiv  S(\sum_c p_c \rho_c) - \sum_c p_c S(\rho_c)$
is the Holevo quantity, which will reappear below in \S\ref{sec:bounds}.  
The Holevo quantity provides a bound on the amount of classical information that can be sent by a quantum channel.
(The discord of $\rho$ itself is defined by extremizing over the choice of measurement on $C$.)

\subsection{Distinguishing QDL and ergodic states with conditional entropy}
\label{subsec:ce-qdl-vs-ergodic}

The conditional entropy is capable of distinguishing between 
QDL and ergodic states.  To do this, we further bipartition each subsystem into spatial regions.
That is, divide the system into $ABCD$ where $A, B$ are light and $C, D$ are heavy.
We assume the total system size is $L = A+B=C+D$.

For the purposes of this argument, to describe an ergodic state, we will use Page's rule \cite{page1993average}:
$$S_\text{ergodic}(A) = \min(A, \bar A) $$
where $A$ denotes the system size, $\dim H_A \sim e^A$.
Therefore
when $A+C$ is less than half the system
\be\label{eq:ergodic-expectation} S_\text{ergodic}(A|C) = S(AC) - S(C) = (A+C) - C  = A \ee
has a volume law in the size of $A$.

To describe a QDL state, 
we will use the model QDL wavefunction on a system of two types of hardcore bosons from \cite{grover2014quantum}:
\be \Psi_\text{QDL}(N,n) = \psi(N) \prod_{j=1}^L {1\over \sqrt{2}} \( \delta_{n_j, 0} + e^{ \ii \pi N_j } \delta_{n_j, 1} \).\ee 
$n_j=0,1$ are the light degrees of freedom and $N_j = 0,1$ are the heavy degrees of freedom.
$\psi(N)$ is an ergodic wavefunction.  When necessary, we assume that it is $\psi(N) = \text{sgn}(\{N_j\})2^{-L/2}$, independently random signs
for each configuration of the heavy degrees of freedom.

First let us determine $S_C$.  The reduced density matrix is
\begin{align} \rho_C(N_C, N_C') & = \sum_{n, N_D} \Psi_{\text{QDL}}(N_C,N_D, n) \Psi_{\text{QDL}}^\star (N_C', N_D,n) 
\\ & = \sum_{N_D} \psi(N_C,N_D) \psi^\star (N_C', N_D) \prod_{j\in C} \( 1 + e^{ \ii \pi (N_j + N_j') } \over 2 \) 
= \delta_{N_C, N_C'} \sum_{N_D} | \psi(N_C, N_D) |^2 . \end{align}
If we use the random-sign form of the wavefunction, this is a diagonal density matrix all of whose eigenvalues are equal
(to $2^{-C}$), so the entropy is maximal.
To estimate the entropy more generally we can compute the purity 
$S_2(\rho_C) \equiv - \log \tr \rho_C^2 \leq S(\rho_C) .$
$$\tr \rho_C^2  = \sum_{N_C,N_C'} \rho(N_C, N_C') \rho(N_C', N_C) = \sum_{N_C} \( \sum_{N_D} | \psi(N_C,N_D)|^2 \)^2
= 2^{- C} .$$
 and therefore
$ S_2(\rho_C)  = C \log 2 $ is volume law in the size of $C$.  Since $S_2 \leq S_{vN}$, this implies that the von Neumann entropy is also volume law.

In the case where $A=C$, the calculation of $S_{AC}$ is done in \cite{grover2014quantum}.  We give the more general calculation 
for $A < C$ because it will be useful below:
\begin{align} \rho_{AC}(N_C, n_A; N_C', n_A') & = 
\sum_{N_D, n_B} \Psi_{\text{QDL}} (N_C,N_D, n_A,n_B) \Psi_{\text{QDL}}(N_C', N_D, n_A', n_B) 
\\ & 
= f(N_C, n_A) f(N_C', n_A') \prod_{j \in C\setminus A} \( 1 + e^{ \ii \pi ( N_j + N_j') } \over 2 \) 
\sum_{N_D} \psi(N_C,N_D) \psi^\star(N_C', N_D) 
\end{align}
where, as in \cite{grover2014quantum}, 
$$ f(N, n) \equiv \prod_{j \in A}  {1\over \sqrt{2}} \( \delta_{n_j, 0} + e^{ \ii \pi N_A} \delta_{n_j, 1} \) .$$
Then
\begin{align} \tr \rho_{AC}^2 
& = \sum_{N_C, N_C'}  
\underbrace{ \sum_{n_A, n_A'} f(N_C,n_A)^2 f(N_C', n_A')^2 }_{= 1} 
\prod_{j \in C\setminus A} \delta_{N_j, N_j'} 
\\ &  ~~~\times ~\sum_{N_D} \psi(N_C, N_D) \psi^\star (N_C', N_D)
\sum_{N_D'} \psi(N_C', N_D') \psi^\star(N_C, N_D') 
 \\ & \simeq \sum_{N_C = N_C'} \sum_{N_D, N_D'} 2^{-2 L }  
 + \sum_{N_D = N_D'} \sum_{N_A, N_A'} \sum_{N_{C\setminus A} } 2^{-2 L }  
 - \sum_{N_D = N_D' = N_C = N_C'}2^{ - 2L } 
 \\& = 2^{ C + 2D - 2 L } + 2^{D + 2 A + (C-A) - 2 L } - 2^{  C+D - 2 L } 
  \\& = 2^{ - C } + 2^{ A-L  } - 2^{  -L  } ~.
\end{align}
Therefore
in this case 
$$ S(AC) = - \log \tr \rho_{AC}^2  = - \log\( 2^{ - C } + 2^{- D - C + A  } - 2^{  -L  } \) .
$$
In the special case $A=C$, this is $S(AC) =  - \log\( 2^{ - C } + 2^{- D  } - 2^{  -L  } \) $ in agreement with
\cite{grover2014quantum}.

Therefore, in a QDL state, we expect
$$ S(A|C)= S(AC) - S(C) = - \log\( 2^{ - C } + 2^{- D - C + A  } - 2^{  -L  } \)  - C \log 2 .$$
Setting $A=C=l$, for $l \ll L$ this behaves as
\be\label{eq:qdl-expectation} S(A|C) \buildrel{l \ll L }\over{\simeq} - 2^{-L}  l + \CO\( 2^{-2L} l^2\) ~~~~~~(A=C=l)\ee
-- a {\it negative} volume law with a coefficient which vanishes exponentially with system size.  

If instead $ A= l$ but we hold fixed $C= L/2$, then
\be\label{eq:Cishalf} S(A|C) \buildrel{l \ll L }\over{\simeq} - 2^{-L/2}  l + \CO\( 2^{-L/2} l^2\) ~~~~~~(A=l, C=L/2)\ee
-- the (negative) volume-law coefficient still vanishes exponentially with system size, but not as fast.

\begin{figure}[h!]
\begin{center}
$$
\parfig{.4}{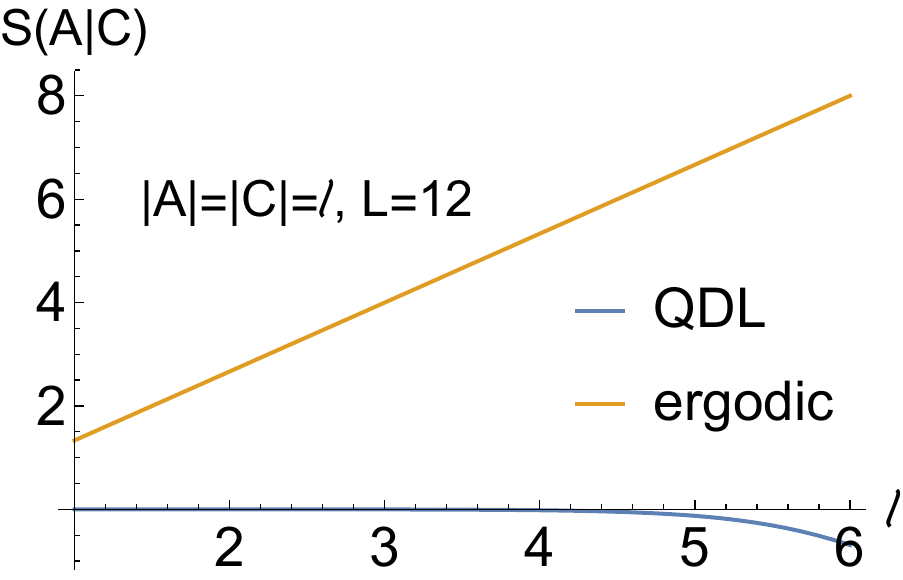} ~~\parfig{.4}{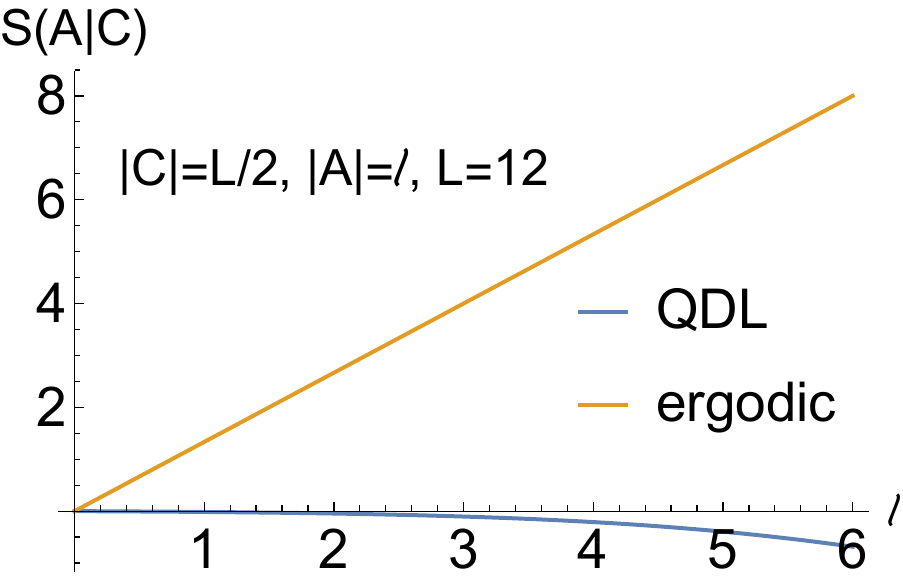}$$
\caption{Expectations for conditional entropy in a QDL state (blue) and an ergodic state (orange), 
based on \eqref{eq:qdl-expectation} and \eqref{eq:Cishalf}.
Left: when regions $A$ and $C$ are the same size, $l$.  Right: when $|C|$ is fixed at $L/2$ and $|A|=l$. 	}
\label{fig:expectation}
\end{center}
\end{figure}

We will compare this expectation with the behavior in scar states of the Hubbard model below in \S\ref{sec:hubbard}.

\subsection{Mutual information QDL diagnostic}

 An alternate protocol replaces the von Neumann entropy 
with the mutual information between $A$ and $B$.
That is, replace the latter two steps of the protocol by
\begin{enumerate}
\setcounter{enumi}{1}
\item  In the resulting state $\rho_{AB}^c$ find the mutual 
information between $A$ and $B$: $I_{\rho_{AB}^c}(A:B)$. 
\item Average over the distribution $p_c$ to obtain an alternate QDL diagnostic
$$ \overline{I_{X_C}(A:B|C)}\equiv \sum_{c} p_c I_{\rho_{AB}^c}(A:B).$$
\end{enumerate}

When $ABC$ is the full system, and measuring $X_C$ completely fixes the state of $C$,
then $\rho_{AB}^c$ is a pure state. To see this explicitly, write the initial state $\ket{\psi_{ABC}}$ 
in the basis of eigenstates of the operator $X_C$: 
$$\ket{\psi_{ABC}} = \sum_c \ket{\psi_{AB}(c)} \otimes \ket{c} .$$ 
Then by the axioms of quantum mechanics, when we measure $X_C$ and 
get the outcome $c$, the resulting state (up to normalization) is
$$ \ket{\psi_{ABC}} \buildrel{\text{measure $X_C$, get $c$} } \over { \to } 
\ket{\psi_{AB}(c)} \otimes \ket{c}, $$
a product state between $AB$ and $C$,
$\rho_{AB}|c = \ket{\psi_{AB}(c)}\bra{\psi_{AB}(c) } $.
In such a state the von Neumann entropy of AB vanishes, $S(\rho_{AB}^c) = 0$, and hence
\be\label{eq:twice} I_{\rho_{AB}^c}(A:B) = S(\rho_{A}^c)+ S(\rho_{B}^c) - S(\rho_{AB}^c)  = 2 S(\rho_{A}^c) .\ee
Averaging \eqref{eq:twice} over the distribution of outcomes $p_c$ says that under these conditions,
the mutual information version of the QDL diagnostic
is twice the original QDL diagnostic:
$$ \overline{I_{X_C}(A:B|C)}  = 2 \overline{S_{X_C}(A) } ~~~~\text{if $ABC$ is pure} .$$

A related quantity which depends on a state $\rho_{ABC}$ but not a choice of operator 
is the conditional mutual information
\begin{align}\label{eq:CMI-def}
 I(A:B|C) & \equiv S_{AC} + S_{BC}  - S_{ACB} - S_C 
= I(A:BC) - I(A:C) 
\\ 
& = D(\rho_{ACB} || \rho_A \otimes \rho_{BC} ) - D(\rho_{AC} || \rho_A \otimes \rho_{C} ) 
\label{eq:CMI-RE}
\end{align}
where $D(\rho||\sigma) \equiv \tr \rho\log \rho - \tr \rho \log \sigma \geq 0 $ is the relative entropy.
In \S\ref{subsec:cmi-qdl} we will bound the alternate QDL diagnostic in terms of the conditional mutual information.

\section{Conditional entropy and QDL physics in the Hubbard model}

\label{sec:hubbard}

To establish the utility of the diagnostics introduced in \S\ref{sec:intro}, 
we compute the conditional entropy and conditional mutual information 
in the same model studied in Ref.~\cite{garrison2017partial}. We perform exact diagonalization of the one dimensional Hubbard model with a repulsive nearest neighbor interaction added to break integrability
\be
\HH = -t\sum_{i\sigma} c^\dagger_{i\sigma} c^\nd_{i+1\sigma} + U \sum_i n_{i\up} n_{i\down} + V \sum_i n_i n_{i+1}
\label{eq:hubbard-ham}
\ee
where $n_{i\sigma} \equiv c^\dagger_{i\sigma} c^\nd_{i\sigma}$ and $n_i \equiv \sum_\sigma n_{i\sigma}$.

\begin{figure}[h!]
 $$
\parfig{.3}{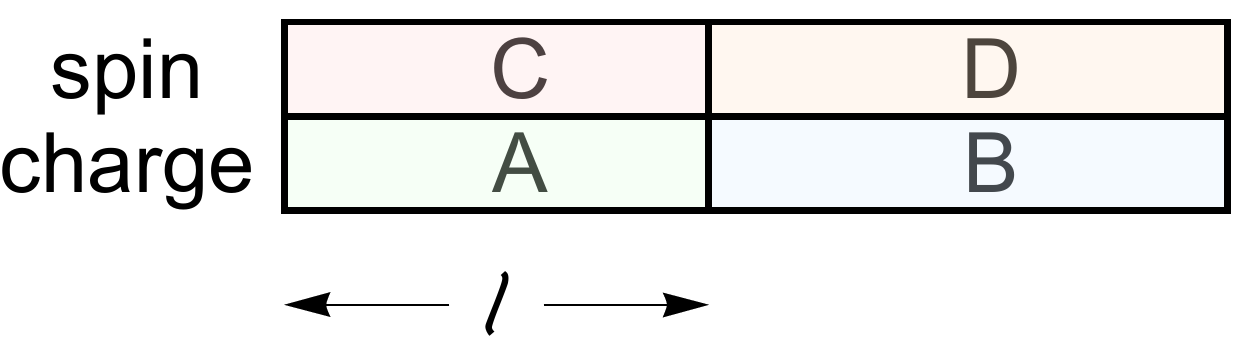}
 \parfig{.7}{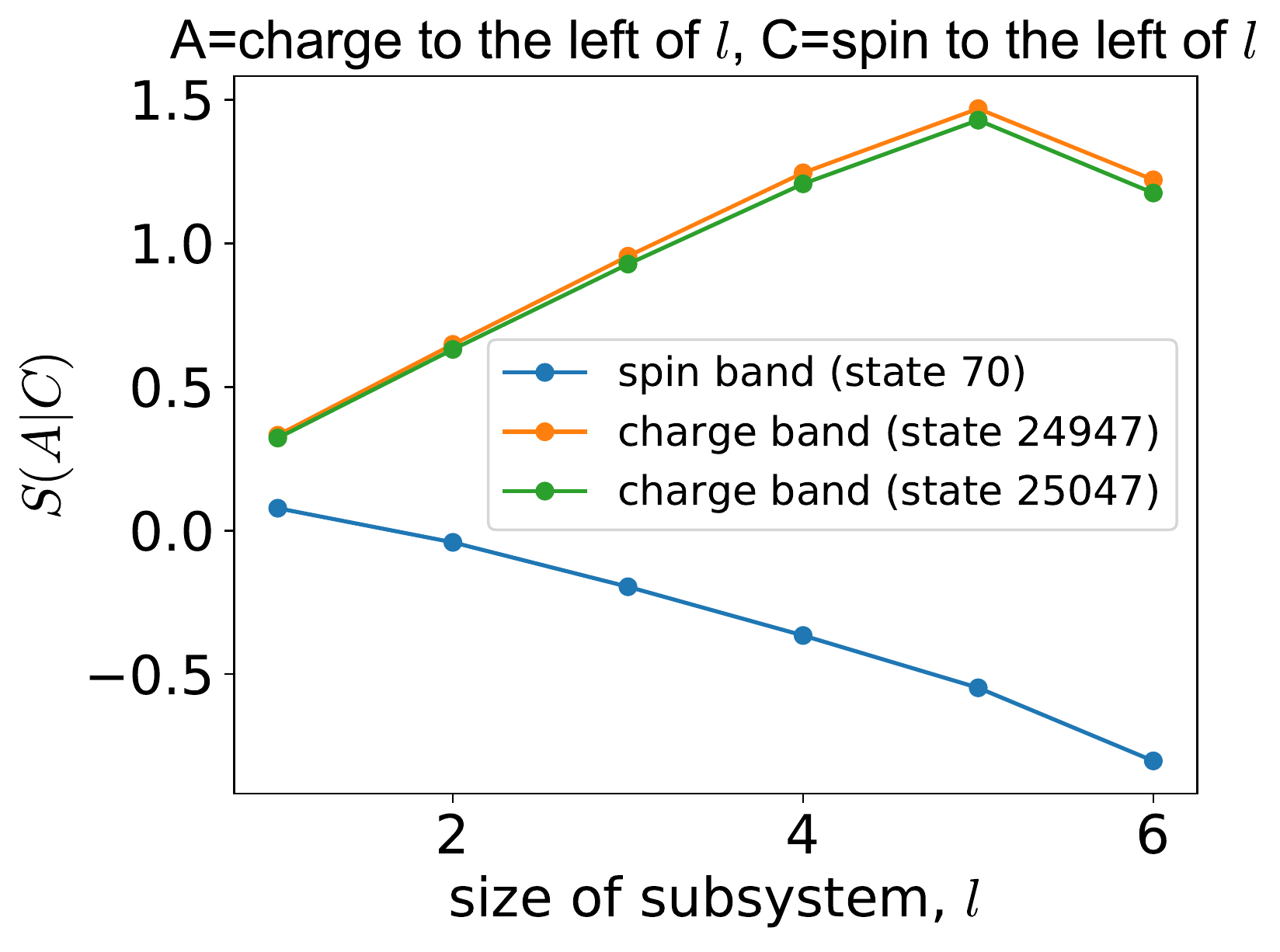}
 $$
 \caption{
 Left: Partition of the Hubbard model degrees of freedom.
  $A$ ($C$) is chosen to be the charge (spin) degrees of freedom on sites $1$ through $l$.
  Right: 
 Conditional entropy in a one dimensional Hubbard model as a function of the location of the bipartioning cut at couplings $U = 4,~V = 3/4$. The chain is periodic and has $L=12$ sites, the particle number is at half-filling, and the magnetization is zero.  
 We show the result for two charge band states and one spin band state, 
 which can be compared, respectively, with
\eqref{eq:ergodic-expectation} and \eqref{eq:qdl-expectation}.
 }
 \label{fig:CE-hubbard}
\end{figure}

Given an eigenstate of $\HH$, we compute the conditional entropy $S(A|C)$
for a subset of the charges (the light degrees of freedom), conditioned on a subset of the spins (the heavy degrees of freedom).
We partition the system as in Fig.~\ref{fig:CE-hubbard}, 
so that subsystem $A$ is all the charge degrees of freedom on sites $1$ through $l$, 
$B$ is the remainder of the charges, 
$C$ is all the spins on sites $1$ through $l$ and $D$ is the remainder of the spins.

In Fig.~\ref{fig:CE-hubbard} we show the result for the conditional entropy $S(A|C) = S(AC) - S(C)$ in several eigenstates.  
Our results confirm those of  \cite{garrison2017partial}, and the striking difference in behavior between ergodic and QDL states illustrates the utility of conditional entropy as a proxy for observing QDL behavior.  
The behavior sharply distinguishes QDL and ergodic behavior, in agreement with our expectations from \S\ref{subsec:ce-qdl-vs-ergodic}.  In particular, in the case of a QDL state, the slope of $S(A|C)$ as a function of the size of $A$ is indeed negative.  
A more quantitative comparison is obstructed by the small system size.

The discussion of the conditional mutual information can be found in Appendix \ref{app:CMI}.

\section{QDL and negativity}

\label{sec:negativity}

For mixed states, the von Neumann entropy is not a good measure of entanglement, since it includes
also classical uncertainty.  A computable measure of entanglement for mixed states is
the logarithmic negativity \cite{eisert99, vidal2002}, defined as 
 $E_N(\rho) \equiv - \log | \rho^{T_A}|  = - \log \( \sum_a |\lambda_a| \) $, where $\lambda_a$ are the eigenvalues of 
 $\rho^{T_A}$, the partially transposed density matrix, 
 $\rho^{T_A}_{ab, a'b'} \equiv \rho_{a'b,ab'}$.

In a QDL state, the reduced density matrix for the `light' particles is essentially separable, because 
$\rho_A = \sum_c p_c \rho_A(c)$ and each $\rho_A(c)$ is area law due to QDL-ness. Therefore the negativity  
 is area law for this mixed state. This result holds irrespective of the size of the Hilbert space of the `heavy' particles, and as we now show, leads to a new diagnostic for the QDL states when the Hilbert space of heavy particles is smaller than half the total Hilbert space.

The area law of negativity is not necessarily a very striking characterization because negativity is area-law even for a Gibbs state 
\cite{sherman2016}.  
Such a thermal state is obtained by tracing out more than half the degrees of freedom in a purely ergodic wavefunction. 
In contrast, for an ergodic wavefunction, integrating out \textit{less than half} the degrees of freedom leads to a \textit{volume law negativity}  \cite{lessThanHalf, bhosale2012}.
The intuition is that nothing dramatic happens if one integrates out a very small region, so negativity will continue to be volume law until the subsystem looks thermal.  This intuition is verified numerically in ergodic spin chains \cite{lessThanHalf} 
and proved analytically for Haar random states \cite{bhosale2012}.

\begin{figure}[h!]
 $$\includegraphics[width=0.7\textwidth]{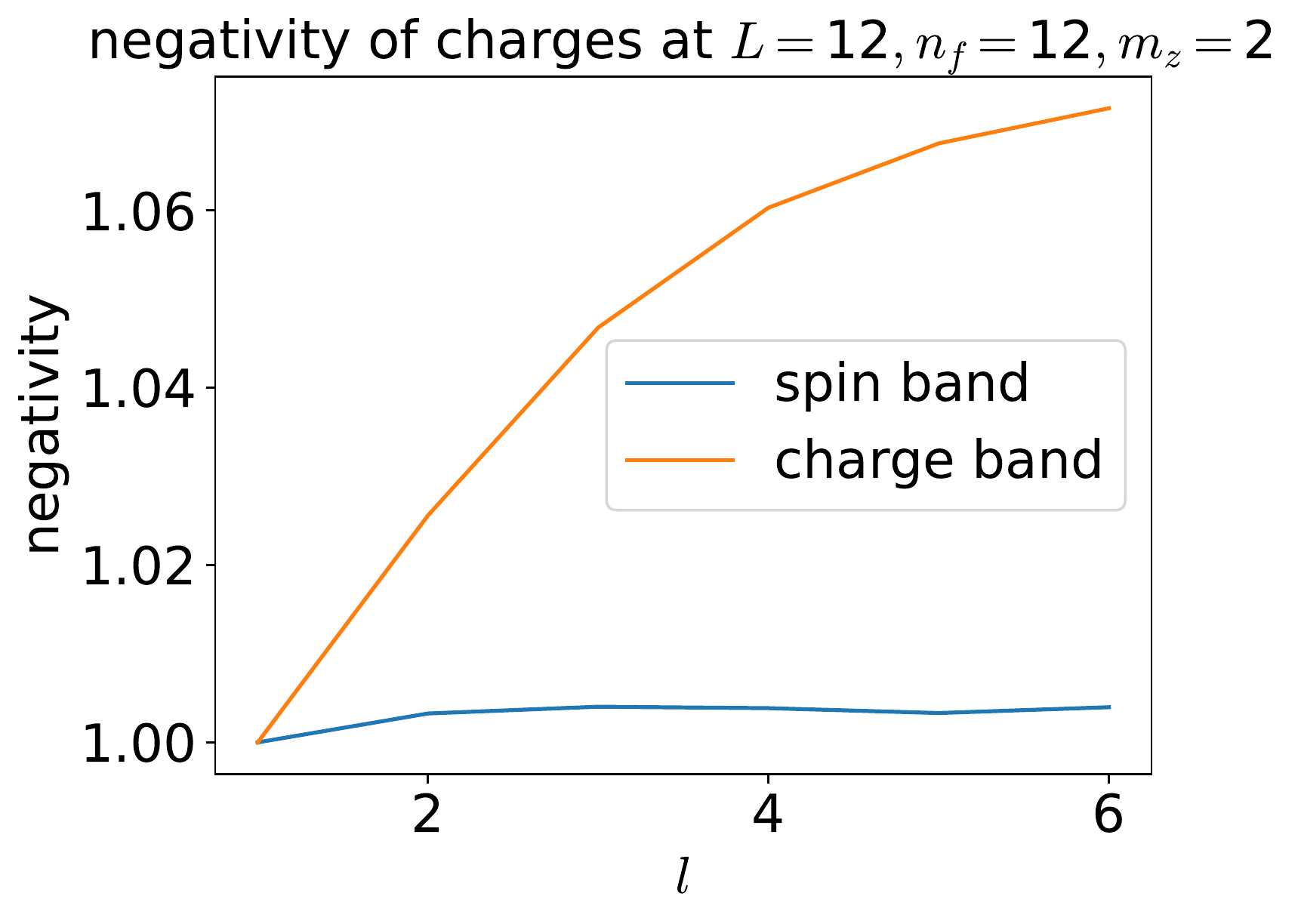}
 $$
 \caption[Logarithmic negativity in QDL and ergodic states]{
 The logarithmic negativity for a spin-band state and a charge-band state of the deformed Hubbard model.  
 Here the magnetization is chosen to be $m_z=2$, at half-filling, so that the spin (heavy) Hilbert space is smaller than the charge (light) Hilbert space.  
 }
 \label{fig:negativity}
\end{figure}

Therefore,  when the Hilbert space of heavy particles is smaller than half the total Hilbert space, then the area law of negativity is another way to quantify QDL behavior.    This expectation is verified in Fig.~\ref{fig:negativity}.

\section{Detecting SPT physics in a gapless system via partial measurement}
\label{sec:hcb}

We now turn to applications of QDL-based protocols to questions about interesting groundstates of 
condensed matter.
We consider a model of spin 1 hardcore bosons governed by the Hamiltonian 
\be
\HH = - t\sum_{i\sigma} b^\dagger_{i\sigma} b^\nd_{i+i\sigma} + h.c. + J \sum_i \vec S_i \cdot \vec S_{i+1} + D \sum (S^z_i)^2.
\label{eq:h_hcb}
\ee
The hopping amplitude $t$ determines an overall energy scale and we set it equal to 1. This model was originally considered in \cite{jiang2018symmetry} as a strong coupling limit of a particular two leg fermionic ladder; the spinful hardcore boson represents two fermions bound into a triplet state across a rung of the ladder. 

In addition to a trivial paramagnet phase, a spin-1 chain also posseses a nontrivial SPT phase known as the Haldane phase \cite{haldane1983continuum,affleck1988valence,kennedy1992hidden}. In Ref.\cite{jiang2018symmetry},
the model in \eqref{eq:h_hcb} was argued to exhibit spin-charge separation in the sense that the ground state wavefunction factorizes into a charge wave function times a spin wave function on the squeezed lattice (the lattice obtained by deleting the unoccupied sites). As a result of this spin-charge separation, it was argued that the spin degrees of freedom can form an SPT phase, despite the presence of the gapless charge degrees of freedom. 

\begin{figure}[h!]
$$ \includegraphics[width=0.9\textwidth]{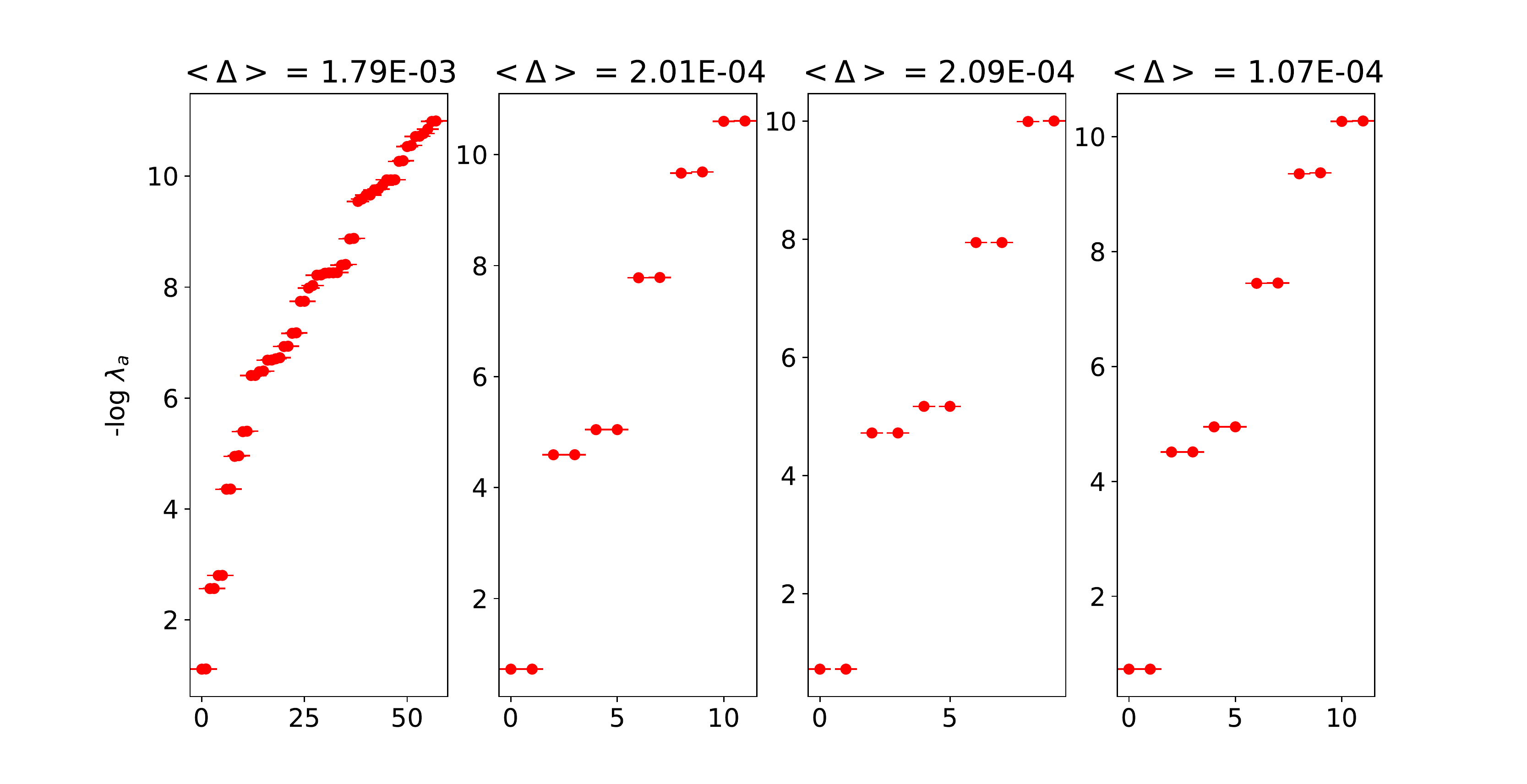}$$
 \caption{Leftmost: Entanglement spectrum across the center bond of the full wave function. Right plots: Entanglement spectrum across the center bond for the projected wave function $\ket{\Psi_s}$ for three randomly chosen hole configurations.  The system size is $L = 96$, with couplings $J = 0.1$ and $D = 0.01$. Also shown is the mean spacing between pair levels in the entanglement spectrum $\vev\Delta$. }
 \label{fig:hcb_entanglement_spectrum}
\end{figure}

Here we apply the QDL protocol to freeze the charge degrees of freedom and study the entanglement properties of spins in the resulting wavefunction. For a conventional spin wavefunction, one signature of an SPT phase is a degeneracy in the entanglement spectrum due to the edge modes \cite{pollmann2010entanglement}. We therefore propose to use the entanglement spectrum of the post-projection wavefunction to access the topological properties of the state.

The ground state of \eqref{eq:h_hcb} was obtained for both open and periodic boundary conditions using DMRG, with bond dimensions up to $\chi = 3000$. Figure \ref{fig:hcb_entanglement_spectrum} shows the post-projection entanglement spectra in the topological phase ($J\gg D$). It is very easy to see the systematic double degeneracy throughout the entire entanglement spectrum. As we increase $D$ and go over to the trivial phase, the degeneracy disappears as expected. 

In addition to the entanglement spectrum, we can look at the entanglement entropy of the post-projection wavefunction. In particular, we study $S_\text{QDL}(c, A)
\equiv \sum_{c} p_c S(\rho^c_A)$ where $p_c$ is the probability of finding a particular configuration $c$ of the (measured) charge degrees of freedom, and $\rho^c_A$ denotes the density matrix corresponding to the (unmeasured) spin degrees of in a subregion $A$.

\begin{figure}[h]
\parfig{.34}{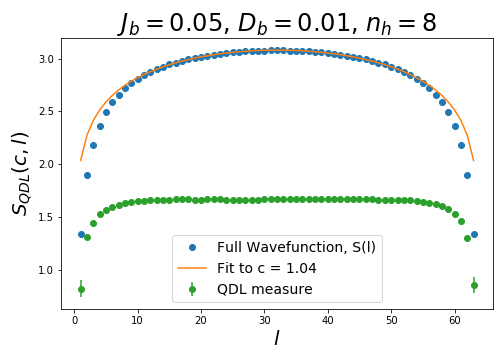}
 \hspace{-20pt} \parfig{.34}{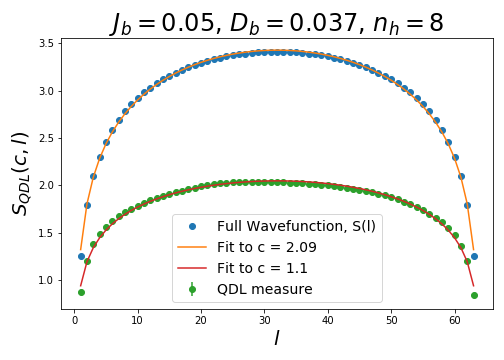}
 \hspace{-20pt}\parfig{.34}{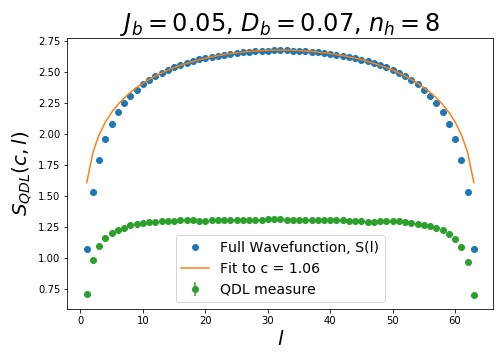}
 \caption{QDL Measure $S_\text{QDL}(c, l)$ across the phase diagram. 
  Left) Topological phase: entanglement entropy of post-projection spin wavefunction collapses to an area law. Middle) Critical point: entanglement entropy of post-projection spin wavefunction shows critical behaviour with central charge $c=1$. Right) Trivial phase: entanglement entropy of post-projection spin wavefunction collapses to an area law. }
 \label{fig:hcb_qdl_ee1}
\end{figure}

We studied \eqref{eq:h_hcb} at system size $L = 64$ with periodic boundary conditions. It is somewhat impractical (and in general impossible) to average over all charge configurations, so we calculated $S_\text{QDL}$ via Monte Carlo sampling with the distribution $p_c$. The results in the topological phase, in the trivial phase, and directly at the critical point are shown in Figure \ref{fig:hcb_qdl_ee1}. In all cases, the entanglement entropy  as a function of subsystem size for the full wavefunction (with no projection) has the form 
\be\label{eq:cft-EE}S(l) \sim \frac{c}{3} \log \left(\frac{L}{\pi} \sin \left(\frac{ \pi l}{L}\right)\right) + b
\ee
with $c = 1$ away from the critical point, and $c=2$ at the critical point. Away from the critical point, the spins are essentially in a gapped phase which mixes very weakly with the gapless charge degrees of freedom, so we see $c=1$ worth of gapless charge.

Applying the QDL projection freezes the charge degrees of freedom, the entanglement entropy of the post-projection wavefunction collapses into an area law behaviour characteristic of a gapped phase in one dimension. 
At the critical point, the post-projection wavefunction retains the entanglement entropy of a critical wavefunction 
\eqref{eq:cft-EE} with precisely $c=1$, in line with the fact that the spins are in a $c = 1$ state at the critical point. 

As far as we are aware, performing partial projections and studying  properties of the `leftover' wavefunction has not been generally explored in the context of ground state wavefunctions\footnote{As we noted above, an interesting exception is
\cite{2013arXiv1307.6617M}, which however does not study gapless states.}. The distinction between an SPT and a trivial phase is typically predicated on the existence of a finite energy gap in both phases \cite{senthil2015symmetry}. However there appear to be examples where features usually associated with SPT phases persist in the presence of gapless modes, including the model studied in this section as well as the models described in \cite{verresen2017topology}. Quantum disentangling is likely to provide a useful method for studying such systems.
 
\section{More bounds on the outcome of the QDL protocol}

\label{sec:bounds}

\subsection{An upper bound on the QDL quantity}

Here we give an upper bound on the QDL quantity in terms of information theoretic objects.
The Holevo bound is a lower bound on the Holevo quantity
\be\label{eq:holevo-def} \chi( \{ p_c , \rho_A^c \} ) \equiv S(\sum_c p_c \rho_A^c)  - \sum_c p_c S(\rho_A^c) \ee
(the entropy of the average density matrix minus the average of the entropies)
in terms of the mutual information between the distribution $p_c$ (call it $X$) and 
that of any measurement $Y$ that can be done on $A$:
$$ 0 \leq H(X:Y) \leq \chi. $$
We use the letter $H$ to emphasize that $H(X:Y)$ is the mutual information between two
classical distributions.
By subtraction, this implies that the QDL quantity is bounded above:
$$ \sum_c p_c S(\rho_A^c) = S(\sum_c p_c \rho_A^c) - \chi \leq S(\sum_c p_c \rho_A^c) - H(X:Y) .$$
The bound gets stronger the bigger is $H(X:Y)$.

The average density matrix is just\footnote{To check this, we can purify $\rho_{ABC}$ by a state $\ket{\psi}_{ABCD}$.  Then
$$ (\rho_A)_{aa'} = \sum_c\sum_{bd} \psi_{abd}^c \psi_{a'bd}^{\star c} = \sum_c p_c (\rho_A^c)_{aa'}. $$} 
$$ \sum_c p_c \rho_A^c = \tr_{BC} \rho_{ABC} = \rho_A.$$
So the first term on the RHS is $S(\sum_c p_c \rho_A^c) = S_A$.
Therefore
\be\label{eq:upper-bound} \QDL \leq S_A - H(X:Y) .\ee
So, at the weakest, we have $ \QDL \leq S_A$ (which follows from concavity of the von Neumann entropy).  

\begin{figure}[h!]
$$ \parfig{.5}{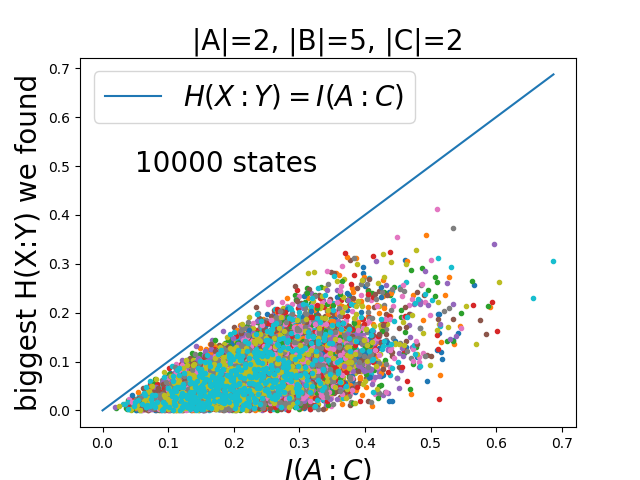}
\parfig{.5}{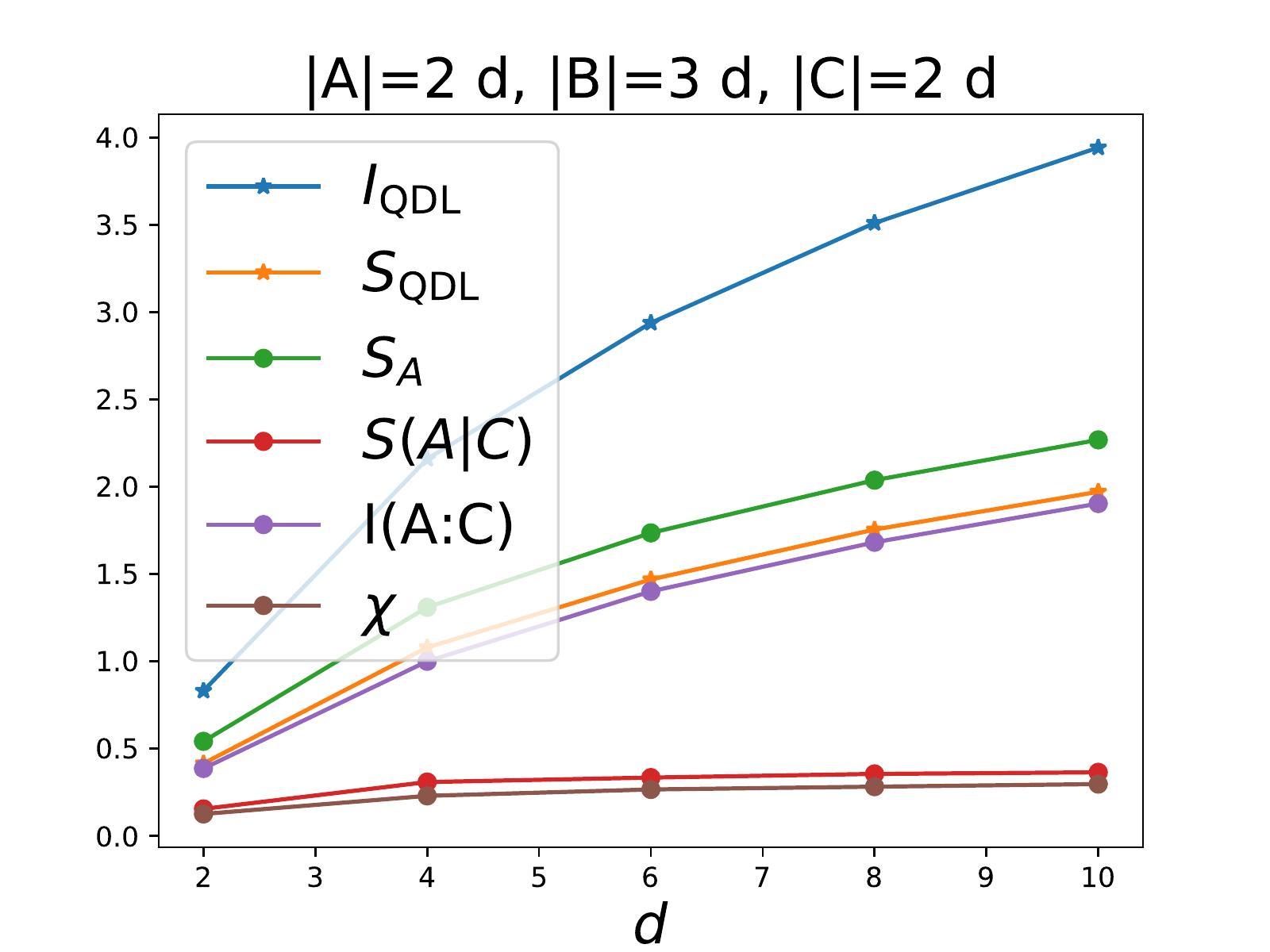}
$$
\caption{Left: Lower bounds on the accessible information between $A$ and $C$ versus 
the mutual information $I(A:C)$ in many Haar-random states on $ABC$ of the indicated dimensions.  
Each dot represents a state of $ABC$; the vertical position is the largest value of $H(X:Y)$, optimizing over measurements $Y$ on $A$. 
We see that there are random states for which the Holevo bound is tight, but on average it is far from being saturated.
Right: various quantities considered in this paper, averaged over 100 Haar-random states on $ABC$, as a function of Hilbert space size, with ratios of dimensions of $A,B,C$ held fixed.  
}
\label{fig:upper-bound-on-QDL}
\end{figure}

To improve upon this estimate, we must ask: as we vary the choice of measurement $Y$, how large can $H(X:Y)$ get?
The largest it can be is called the {\it accessible information}
\be \CI \equiv \max_Y H(X:Y) .\ee
We note that 
\be\label{ineq:accessible} H(X:Y) \leq I(A:C) .\ee 
This follows from \eqref{ineq:MRE} because the distribution $p_{xy} \ket{x}\otimes\ket{y} $ is the 
state that results from measuring the operators $X,Y$ on $A \otimes C$ and not looking at the answer. 
A lower bound on the accessible information just in terms of $ \rho_A = \sum_c p_c \rho_A^c $
is given in \cite{PhysRevA.49.668}, but is not useful for our purposes, because the bound (the ``subentropy" of $\rho_A$) is itself a bounded quantity, independent of the size of $\CH_A$.
In Haar random states, numerical experiments (Fig.~\ref{fig:upper-bound-on-QDL}) show that the inequality \eqref{ineq:accessible} is far from saturated on average.

\subsection{QDL mutual information and CMI}

\label{subsec:cmi-qdl}

Denoting the QDL mutual information 
$$  \overline{I_{X_C}(A:B|C)} \equiv \sum_{c} p_c I_{\rho_{AB}^c}(A:B) \equiv  \IQDL ,$$
we will show that it can be bounded above and below in terms of the conditional mutual information: 
\be\label{ineq:both} I_\rho(A:B|C)  - (I_\rho(AB:C) - I_{\CE \rho}(AB:C) ) \leq \IQDL \leq 
I_\rho(A:B|C)  + (I_\rho(A:C) - I_{\CE \rho}(A:C) ) . \ee
Here $\CE$ is the quantum channel \eqref{eq:diagonal-image} associated with the measurement $X_C$.
The quantity appearing in the error terms 
$ I_{\CE \rho}(A:C) = \chi(\rho^c_{A}, p_c)$
is again the Holevo quantity.  Moreover, 
in both the lower and upper bound, 
$I_\rho(A:C) - I_{\CE \rho}(A:C)$
is again the quantum discord.  

To see the upper bound on $\IQDL$, use \eqref{ineq:MRE} in the first term of 
\begin{align}
I_{\rho}(A:B|C)  & = D(\rho_{ABC} || \rho_{A}\otimes \rho_{BC} ) -D( \rho_{AC}|\rho_A \otimes \rho_C ) 
\\ & \geq D(\CE \rho_{ABC} || \CE( \rho_{A}\otimes \rho_{BC}) ) -D( \rho_{AC}|\rho_A \otimes \rho_C ) 
\\ & = S_{\CE\rho}(A) + S_{\CE\rho}(BC)-  S_{\CE\rho}(ABC) - I_\rho(A:C) .
\end{align}
In terms of the spectral decomposition
of $\rho_{AB}^c = \sum_i \lambda^{(c)}_i \ketbra{\lambda^{(c)}_i}{\lambda^{(c)}_i}$, 
in the state \eqref{eq:diagonal-image}
we have
$$S_{\CE(\rho)}(ABC)  =\sum_c p_c  \sum_i \lambda^{(c)}_i \log p_c \lambda^{(c)}_i = 
H(p) + \sum_c p_c S(\rho_{AB}^c) $$
and similarly for $S_{\CE(\rho)}(BC)$.
In contrast, $S_{\CE\rho}(A) = S\( \sum_c p_c \rho_A^c \) $.
Use \eqref{ineq:MRE} in the first term of 
\begin{align}
I_{\rho}(A:B|C)  & \geq 
S( \sum_c p_c \rho_A^c )
+ H(p) + \sum_c p_c S\( \rho_B^c\) 
- \( H(p) + \sum_c p_c S\( \rho_{AB}^c\) \) 
- I_\rho(A:C) 
\\ &  = \sum_c p_c \( S( \rho_B^c)  
{\color{blue} + S\( \rho_{A}^c\)  }-  S\( \rho_{AB}^c\) \) 
{\color{blue} - \sum_c p_c  S\( \rho_{A}^c\)  }
+ S( \sum_c p_c \rho_A^c ) - I_\rho(A:C) 
\\ & = 
\sum_c p_c I_{\rho^c_{AB}}(A:B)
+ S( \sum_c p_c \rho_A^c )  - \sum_c p_c  S\( \rho_{A}^c\) 
- I_\rho(A:C) 
\\& =  \overline{I_{\CO_C}(A|C)} 
+ \chi\( \{p_c, \rho_A^c\} \)  - I_\rho(A:C) 
\\& =  \overline{I_{\CO_C}(A|C)} 
+ I_{\CE\rho}(A:C) - I_\rho(A:C) ~~.
\label{eq:bound1}
\end{align}
(Indicated in blue are terms which are added and subtracted.)

In the other direction, we can bound $I(A:B|C)$ from above in terms of the QDL mutual information:
\begin{align}
I_\rho(A:B|C) & = S_\rho(A|C) + S_\rho(B|C) - S_\rho(AB|C)  
\\& = - D(\rho_{AC} || \Ione_A \otimes\rho_C ) 
- D(\rho_{BC} || \Ione_B \otimes\rho_C ) 
- S_\rho(AB|C)  
\\& 
\buildrel{\eqref{ineq:MRE}}\over{\leq }
- D(\CE\rho_{AC} || \CE\Ione_A \otimes\rho_C ) 
- D(\CE\rho_{BC} || \CE\Ione_A \otimes\rho_C ) 
- S_\rho(AB|C)  
\\& 
= S_{\CE \rho}(AC) - S_{\CE\rho}(C)
+ S_{\CE \rho}(BC) - S_{\CE\rho}(C)
- S_\rho(AB|C)  
\\& = \sum_c p_c \( S_{\rho^c}(A) + S_{\rho^c}(B) \)  - S_\rho(ABC) + S_\rho(C)
\\& = \sum_c p_c I_{\rho^c}(A:B) + \sum_c p_c S_{\rho^c}(AB)+ S_\rho(C)- S_\rho(ABC)
\\ & = \IQDL + \sum_c p_c S_{\rho^c}(AB)   + S_\rho(C)- S_\rho(ABC)
\\ & = \IQDL + \sum_c p_c S_{\rho^c}(AB)  {\color{blue} - S_\rho(AB) + S_\rho(AB) }  + S_\rho(C)- S_\rho(ABC)
\\ & = \IQDL - \chi(p_c, \rho_{AB}^c)  + I_\rho(AB:C)~.
\label{eq:upper-bound-of-CMI}
\end{align}

Note that 
$$ S_{\CE \rho}(A|C) = \sum_c p_c S_{\rho^c}(A) + H(X)  - H(X)  = \sum_c p_c S_{\rho^c}(A). $$

We have not yet found an example where the error terms in the bound \eqref{ineq:both} 
do not scale with system size.  Numerical experiments (Fig.~\ref{fig:IQDL-CMI}) on random states show that 
while neither directly bounds the other, 
the IQDL quantity and the CMI exhibit the same scaling behavior.
\begin{figure}[h!]
$$ 
\parfig{.33}{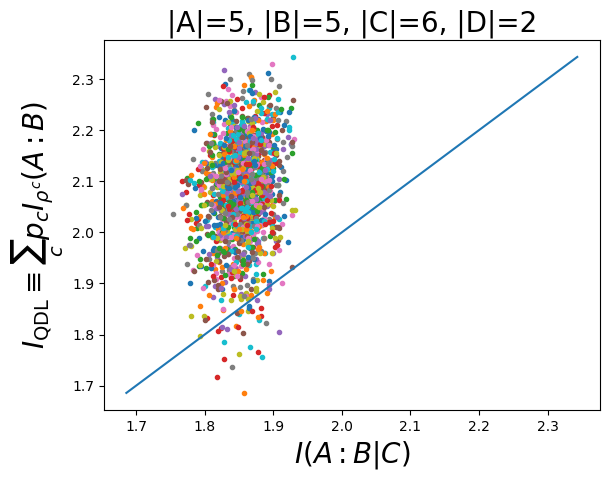}
\parfig{.33}{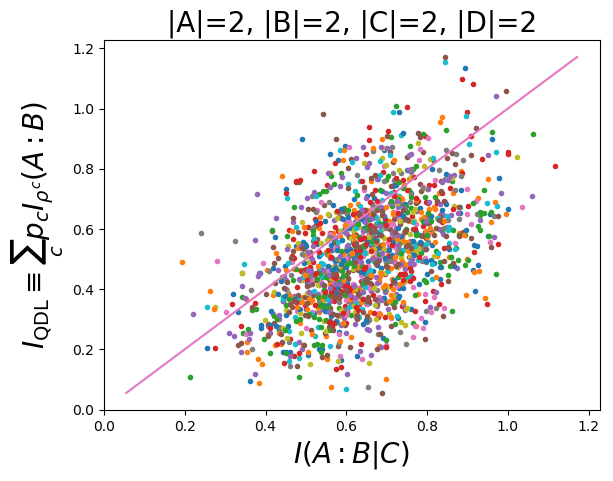}
\parfig{.33}{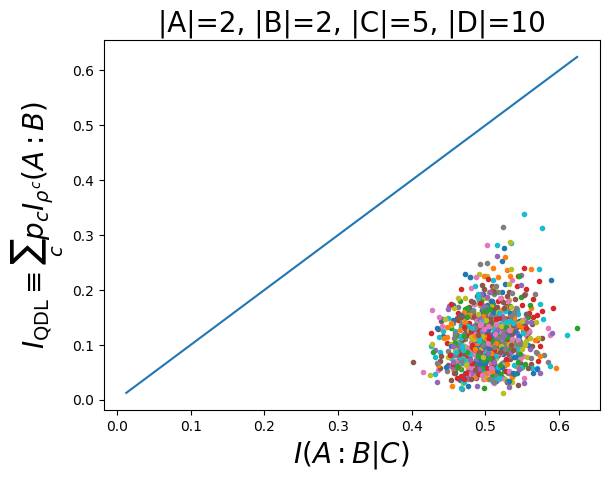}
$$
$$
\parfig{.6}{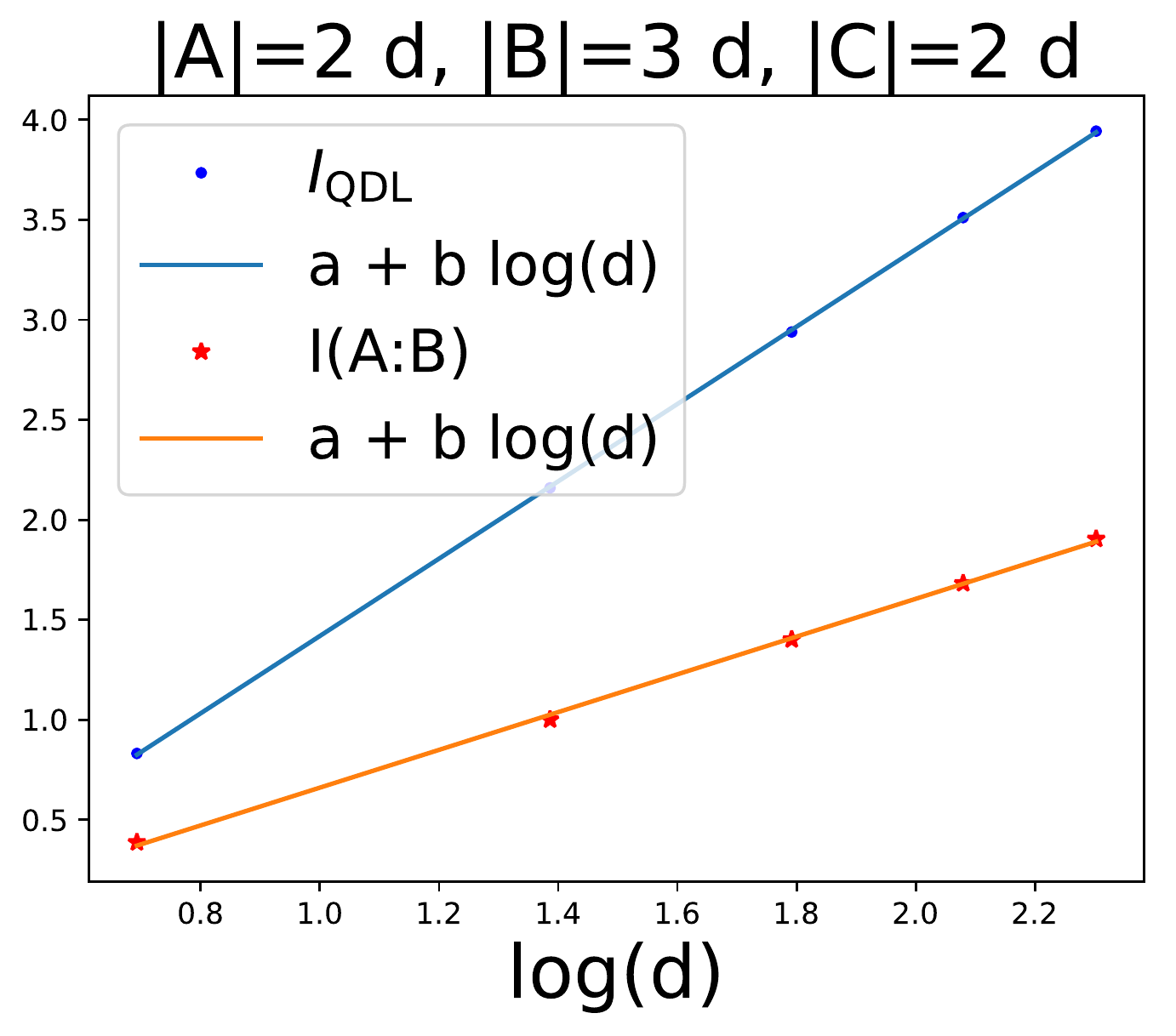} $$
\caption{Top: the $I_\text{QDL}$ quantity versus $I(A:B|C)$ for many pure Haar-random states of $ABCD$, with $ABCD$ of the given dimensions.  Bottom: a fit to the average behavior of both the $I_\text{QDL}$ quantity and the CMI in Haar-random states as a function of the overall Hilbert space size $d$, holding fixed the ratios of dimensions of $ABC$ ($|D|=1$ here).  Both exhibit a volume law in system size $\propto \log d$.
}
\label{fig:IQDL-CMI}
\end{figure}

\section{Disentangling heavy Fermi liquids}

\label{sec:heavy-fermions}

We now turn to the applications to heavy-fermion physics. An elementary model for heavy fermion materials 
is the Kondo lattice model (KLM).  The KLM consists of a lattice of localized moments $\vec S$ coupled to a sea of conduction electrons by spin exchange.
\be\HH_{K} =  \sum_{k}\epsilon(k) c^\dagger_k c^\nd_k + J_K \sum_i \vec S_i \cdot \vec s_i + J_H \sum_{\vev{ij}} \vec S_i \cdot \vec S_j\ee
where $\vec s = \tfrac{1}{2} c^\dagger_\alpha \vec \sigma_{\alpha\beta} c^\nd_\beta$ is the electron spin. We have also included the possibility of antiferromagnetic exchange interactions between the local spins. This model has been extensively studied and exhibits several phases. 
The phase diagram is determined by a competition between an RKKY effect which favors a magnetically ordered state, and the Kondo interaction. 
In addition to the Heavy Fermi Liquid (HFL) phase which has a large Fermi surface (FS), there is possibility of an alternative paramagnetic state where the spins decouple from the conduction electrons and enter a spin-liquid state  \cite{FLstar}. Such an `FL$^*$ phase' is characterized by a fractionalized spin liquid coexisting with a small FS of conduction electrons.

One important distinction between the HFL phase and the FL$^*$ phase is that in the former, the conduction electrons and the local moments are entangled at long distances, while in the latter, they are not. One way to characterize this entanglement is to consider the mutual information between these two degrees of freedom in a given spatial region, as discussed in Refs.\cite{hofmann2019, toldin2019}. Within this scheme, one considers local moments $f_A$ and conduction electrons $c_A$ in a given subregion $A$, and considers $S(f_A) + S(c_A) - S(f_A \cup c_A)$. One potential drawback of this quantity is that  both $S(f_A)$ and $S(c_A)$ are sensitive to short distance entanglement between local moments and conduction electrons in region $A$, and will generically be volume law. This short distance volume law entanglement is not canceled out by the subtracted term $S(f_A \cup c_A)$ which satisfies an area-law upto multiplicative logarithmic corrections. Here we instead explore QDL inspired ideas to study the nature of entanglement between the spins and the conduction electrons. Compared to a mutual-information-based protocol,  we will find that a QDL based protocol can be devised which is sensitive only to long distance entanglement.

In a system with a FS, the entanglement entropy of a region of linear size $l$ will behave as 
$S(l) \sim l^{d-1} \log l$, a violation of the area law \cite{gioev2006, wolf2006}.
Consider measuring the positions of all conduction electrons in the ground state of the KLM.  In an HFL state, the local moments participate in the Fermi surface, and we expect that the resulting wavefunction will continue to have the properties of a state with a Fermi surface, namely, $S(l) \sim l^{d-1} \log l$.  In contrast, in the FL$^*$ state where the local moments form a gapped spin-liquid, one expects that  the resulting wavefunction satisfies $S(l) \sim l^{d-1} - \gamma$ where $\gamma$ is the topological entanglement entropy corresponding to the topological order of the local moments. This is analogous to our discussion of gapless SPT state in Sec.~\ref{sec:hcb}, where the charge degrees of freedom form a Luttinger liquid, while the spin degrees of freedom form an SPT state.

Motivated by the discussion in \S\ref{sec:bounds},
we consider the conditional mutual information between two non-overlapping sets of local moments $A$ and $B$, 
conditioned on the state of all the conduction electrons, $C$.
\be I(A:B|C) = S_{AC} + S_{BC} - S_{ABC} - S_C = S_A + S_B - S_{AB}  = I(A:B).
\ee
where the last relation follows because $ABC$ is the full system.  
We consider the system on a torus of dimensions $L_x \times L_y$, and 
cut the system at fixed $x$ into two cylinders $A$ and $B$.
Due to the local Kondo hybridization, the entanglement entropy of local moments will be dominated by a volume law. Schematically, we expect the entanglement of spins in a region of size $L_A$ (large compared to the lattice spacing) to be described by 
\be S^{\text{local~moments}}(l)= a_1 L_y L_A+  a_2 L_y \log \text{min}(L_A, L_B)  + \cdots. \label{eq:s_local_mom}\ee
In the HFL phase, spins participate in the Fermi surface and we expect the area-law violating coefficient $a_2$ to be nonzero. 
The dependence on $L_y$ and $L_A$ may be understood by thinking of the 2d system on the torus as a collection of wires running in the $x$ direction, one for each value of the conserved momentum $k_y$ \cite{Swingle:2009bf,Swingle:2010yi}.

\begin{figure}[h!]
$$ \parfig{.5}{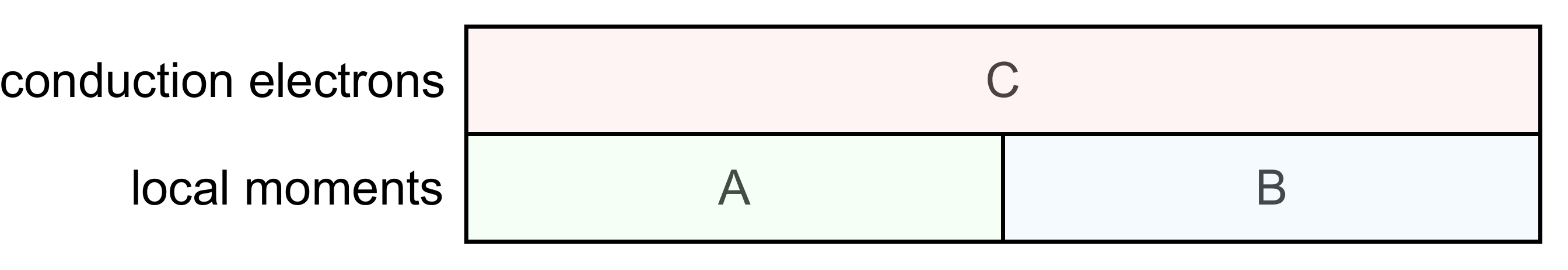}$$
\caption{Subsystems involved in the conditional mutual information which exposes the Fermi surface behavior of the local moments.}
 \label{fig:hfl-regions}
\end{figure}

The mutual information between spins in $A$ and spins in $B$ conditioned on all itinerant electron degrees of freedom provides a subtraction scheme for subtracting out the volume law contribution in Eq.~\ref{eq:s_local_mom}, thereby exposing the coefficient $a_2$. 
With the definitions in Fig.~\ref{fig:hfl-regions}, for the case $l = L/2$ where $A$ and $B$ are each half of the spins, we have
\be I(A:B|C) = S_A + S_B - S_{AB} 
\sim 2 a_2 L_y \log L_A . \ee

\begin{figure}[h!]
 $$\includegraphics[width=0.5\textwidth]{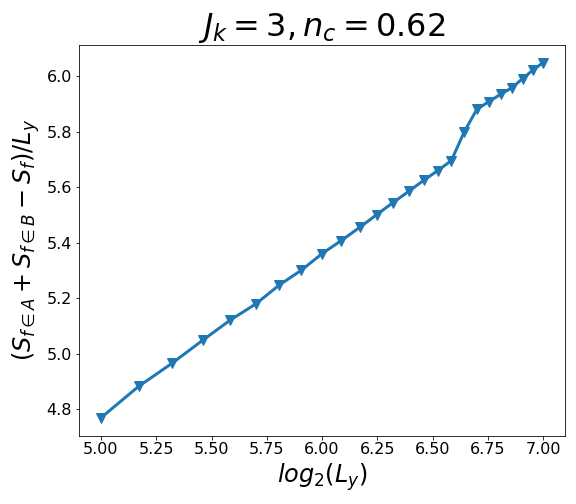} \includegraphics[width=0.5\textwidth]{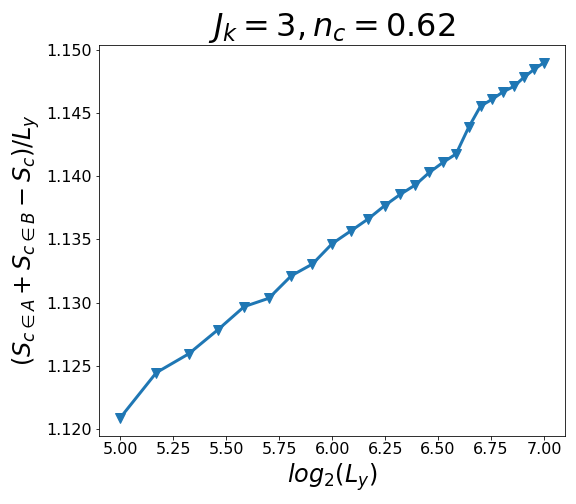}$$
 \caption{Conditional mutual information calculated in the paramagnetic HFL phase, in mean field theory. The sample
has dimensions $2L_y \times L_y$.
 It is bipartitioned into left and right $L_y \times L_y$ halves, $A$ and $B$, as in Fig.~\ref{fig:hfl-regions}; we vary the full system size $L_y$.  Left: Conditional mutual information of $f$ electrons as a function of system size, $I(f\in A: f\in B| c) = I(f\in A: f\in B)$ . Right: Conditional mutual information of $c$ electrons, $I(c\in A:c\in B| f) = I(c\in A: c\in B)$. }
 \label{fig:hfl-cmi}
\end{figure}

\begin{figure}[h!]
 $$\includegraphics[width=0.5\textwidth]{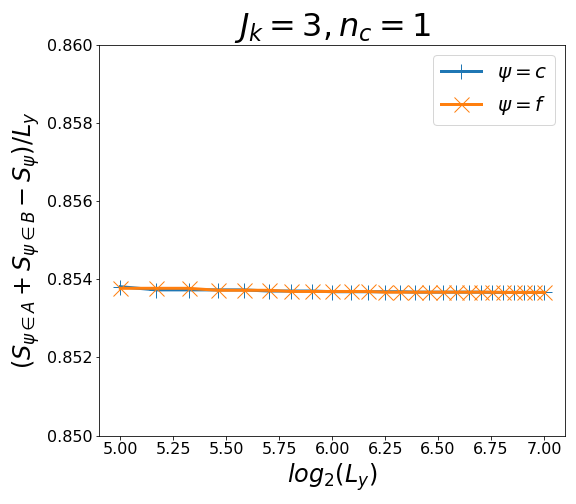}
 \includegraphics[width=0.5\textwidth]{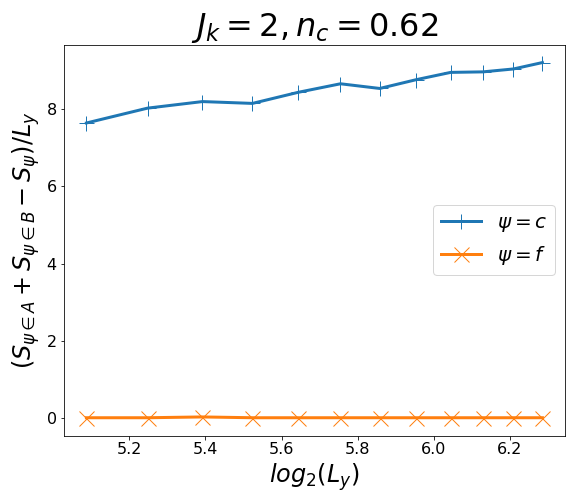} $$
 \caption{In these plots we show conditional mutual information in different parts of the phase diagram, within mean field theory. 
Each plot shows both $S_{f\in A} + S_{f\in B } - S_f$ (crosses) and $S_{c\in A} + S_{c\in B } - S_c$ (vertical lines),
in the geometry described in Fig.~\ref{fig:hfl-regions},
as a function of overall system size $L_y$.
 Left: Conditional mutual information at a value of couplings in the HFL phase but at $n_c = 1$.  Both 
 curves exhibit a strict area law indicative of no Fermi surface. Right:  Conditional mutual information calculated in the AFs phase.  The conduction electrons show behavior indicative of a Fermi surface, while the local moment fermions do not.}
 \label{fig:afs-cmi}
\end{figure}

Here we restrict ourselves to  a mean-field treatment of the KLM where the local moments are represented by Abrikosov/slave fermions $\vec S_i = f^\dagger_{i\alpha} \vec\sigma_{\alpha\beta} f^\nd_{i\beta}$ \cite{read1983, read1984}. When the Kondo coupling is sufficiently large, one obtains the HFL phase which is characterized by a mean field order parameter $V = \vev{c^\dagger f}$ representing the hybridization between $c$ and $f$ electrons.

Since the mean-field Hamiltonian is quadratic in fermion creation/annihilation operators, we can calculate the entanglement entropy of subsets of degrees of freedom using the correlation matrix technique \cite{peschel2003calculation}. Specifically, we consider a rectangular system of size $2L_y \times L_y$, and measure the mutual information between f-electrons (or c-electrons) in the left and right halves of the system. The results in the HFL phase are shown in Fig. \ref{fig:hfl-cmi}. Entanglement entropy of $f$ or $c$ electrons alone follows a volume law.  In contrast, their mutual information behaves as $I(A:B) \sim l \log l$, providing evidence that both $f$ and $c$ participate in the Fermi surface. We also studied the special case of $n_c = 1$, which corresponds to a Kondo insulator, and we find that the mutual information of both $f$ and $c$ saturates to an area law without any multiplicative logarithmic correction, as expected (Fig.~\ref{fig:afs-cmi}, left).

When the Kondo screening is not operative, the mean field description of the system has $V = 0$. As mentioned above, a natural state with $V = 0$ is the  FL$^*$ state. An alternative possibility that exists within mean-field is an
antiferromagnetic ordered state of local moments decoupled from the conduction electrons \cite{ogata2007, becca2013}. The right panel of Figure \ref{fig:afs-cmi} shows that, as expected, the mutual information of $c$ fermions is Fermi-surface-like, while the $f$ fermions are in a product state with no entanglement (an `AFs' state  \cite{ogata2007, becca2013} where `s' denotes small Fermi surface).

The mean field parameter $V$ is not gauge invariant, and therefore it vanishes if the constraint $n_f = 1$ is implemented exactly. We propose that the conditional mutual information provides a gauge-invariant order parameter for the HFL phase, and the  FL$^*$ state. Although it is difficult to imagine measuring the mutual information experimentally for a macroscopic quantum system, it would be extremely interesting to implement our scheme within a Gutzwiller projected wavefunction where $n_f = 1$ is satisfied exactly on each site. We leave this for future work.


\section{Discussion}

	In this paper we generalized and employed the idea of quantum disentangled liquids (QDL) introduced in Ref.~\cite{grover2014quantum} in several different directions: \begin{enumerate}
		\item  We obtained a relation between the QDL quantity  and conditional entropy, which provides an operator-agnostic definition of a QDL phase. In particular, we showed that in a finite energy eigenstate belonging to a QDL state, conditional entropy of light degrees of freedom is negative with a vanishingly small volume law coefficient, in contrast to an ergodic state, where it is positive, with an $O(1)$ volume law coefficient.  In addition, we  showed that the scaling of entanglement negativity can also sharply distinguish between a QDL state and an ergodic state.
		
		\item We argued that a QDL-based protocol can detect topological invariants in a gapless topological phases by studying a concrete model where charge degrees of freedom form a Luttinger liquid while the spin degrees of freedom are in an SPT state. 
		\item We argued that a QDL-based protocol can be used to detect universal features of entanglement in Kondo lattice systems and can serve as an order parameter for a heavy Fermi liquid. 
	\end{enumerate}
		Further, we obtained several inequalities relating the QDL quantity to conditional information theoretic quantities.

Broadly speaking, our approach provides a new way to characterize entanglement in multi-component systems. It stands in contrast to more commonly used field-space entanglement \cite{Taylor:2015kda, Mozaffar:2015bda,Mollabashi:2014qfa}, or  particle-space entanglement \cite{zozulya2007, herdman2014}, both of which lead to volume law entanglement even in the ground state of a local Hamiltonian due to non-local bipartitions of the Hilbert space. This makes it harder to separate universal contributions to entanglement entropy. In contrast, the measurement-based QDL quantity as well as conditional entropy follow an area law in the ground state (upto logarithmic corrections).

We derived several inequalities relating QDL quantity to operator-agnostic measures such as conditional entropy. These inequalities can be thought of as a manifestation
of the monogamy of entanglement, which simply states that  if party $A$ is strongly entangled with party $B$, then it can't entangle strongly with another party $C$. For example, in the QDL phase, the light degrees of freedom with Hilbert space $A$ are strongly entangled only with the heavy degrees of freedom in their immediate vicinity, which we denote as $C$. Therefore, measuring $C$ disentangles $A$ leading to an area-law scaling for the QDL quantity. Monogamy of entanglement implies that $A$ would be unable to entangle with heavy degrees of freedom which are not in their immediate vicinity, and therefore, $S(AC) \approx S(C)$, leading to a small value for the conditional entropy $S(A|C)$.

Gapless topological phases are poorly understood in general. Examples in one dimension include \cite{verresen2017topology, 2019JonesVerresen, Verresen:2019igf} as well as the model in Eq.~\eqref{eq:h_hcb}  from Ref.\cite{jiang2018symmetry}, the subject of our discussion in Sec.~\ref{sec:hcb}. As briefly discussed in Sec.~\ref{sec:heavy-fermions}, we expect that in a fractionalized Fermi liquid (the FL${}^{*}$ phase) of Kondo lattice model, the conditional mutual information of local moments will satisfy an area law, and also contain a subleading non-zero topological entanglement entropy (assuming that the local moments are in a gapped topological state). It will be interesting to extend this idea to gapless phases obtained via slave particle construction. Consider, for example, the Halperin-Lee-Read (HLR) state \cite{HL9312}, which is a compressible quantum Hall state found in the half-filled Landau level. This phase can be understood in terms of a parton construction
\cite{2012PhRvB..86g5136B}, $ c = f b $ 
where $c$ is the annihilation operator for the electron, the fermion $f$ forms a Fermi liquid and $b$ forms an incompressible fractional quantum Hall state at $\nu=\half$. The identification of diagnostics which reveal the topological order hidden in this gapless state is a long-standing problem \cite{gaplessTO}. Might one be able to reveal it using the ideas described in this paper? For example, can one devise procedure which projects out only the degrees of freedom corresponding to fermion $f$? A related question arises in the context of spin-3/2 spin chain, where it has been argued that despite the gapless spin degrees of freedom in the bulk, there still exist topological edge states \cite{ng1994}. Can one project out the effective spin-1/2 degrees of freedom to reveal the edge states corresponding to the effective spin-1 degrees of freedom?

A comment is in order about the use of measurement-based protocols for groundstate properties.
One may have the impression that the projection onto the measurement outcome is a very violent operation. 
This makes it not obvious that the post-measurement wavefunction is still sensitive to subtle
low-energy properties of the original state.  However, a representation of the resulting amplitudes 
in terms of a path integral makes clear that the measurement projection only changes the 
{\it boundary conditions} on the path integral (at the Euclidean time slice where the wavefunction is evaluated),
and therefore does not change its universal properties.

Finally, it will be interesting to consider implementing the measurement of QDL quantity  in experiments  to put bounds on conditional entropy using Eq.\ref{eq:ce-bound}. Naively, when $|A| = O(1)$ and $|C| \gg 1$, one might think that QDL quantity can be measured without much difficulty by performing a projecting measurement on $C$ followed by a state tomography on $A$. However, a major challenge with this approach is that state tomography requires an $O(|A|)$ destructive measurements on $A$, and the outcome of projective measurement on $C $ prior to these destructive measurements should be \textit{identical}. This is because the QDL quantity involves  $S(\rho^c_A)$, the density matrix $\rho^c_A$ on $A$ for a fixed outcome $c$ in $C$. This will be challenging when $|C| \gg 1$. Despite these difficulties, QDL quantity is easier to measure than the conditional entropy because the latter will require state tomography on $C$ as well.

\vfill\eject
\noindent
{\bf Acknowledgements}

   We acknowledge useful discussions with and comments from Emily Davis,
Tsung-Cheng Peter Lu, Alex Meill, Max Metlitski, Masaki Oshikawa and Ashvin Vishwanath.
  This work was supported in part by funds provided by the U.S.\
  Department of Energy (D.O.E.)\ under cooperative research agreement
  DE-SC0009919.  The work of JM is supported in part by the 
  Simons Collaboration on Ultra-Quantum Matter. TG is supported
  by the National Science Foundation under Grant No. DMR-1752417, and as an Alfred P. Sloan Research Fellow.

\appendix

\section{Details of the lower bound on the QDL diagnostic}

\label{app:bound-details}

Here we give the full details of the proof that $S(A|C) \leq \QDL$.  
\begin{align}
S(A|C)
& \buildrel{\eqref{ineq:MRE}}\over{\leq} A - D(\CE_C(\rho_{AC}) || \CE_C(u_A \otimes \rho_C) ) 
\\ & =A  - \tr \sum_c p_c \rho_A^c \otimes \ketbra{c}{c} \log \( \sum_{c'}p_{c'} \rho_A^{c'} \otimes\ketbra{c'}{c'} \) 
+   \tr \sum_c p_c \rho_A^c \otimes \ketbra{c}{c} \log \( \sum_{c'}p_{c'} u_A \otimes\ketbra{c'}{c'} \) 
\\ & = A - \sum_c p_c  \tr_A \rho_A^c  \bra{c} \( \log \( \sum_{c'}p_{c'} \rho_A^{c'} \otimes\ketbra{c'}{c'} \) 
- \log \( \sum_{c'}p_{c'} u_A \otimes\ketbra{c'}{c'} \)  \) \ket{c} 
\\ & \equiv A + A_1 + A_2 .
\label{eq:wingedvictorymotherfucker}
\end{align}
To evaluate $A_{1,2}$ we must find the eigenbasis of the operators inside the log.  
The eigenvectors of $u_A \otimes \sum_{c'} p_{c'}\ketbra{c'}{c'} $ are $\ket{a}_A\otimes \ket{c}_C $ 
(where $\{ \ket{a}\}$ is any basis for $A$)
and the eigenvalues are $ p_{c'} e^{-A}$.  Therefore
\begin{align} A_2 & = + \sum_c p_c  \tr_A \rho_A^c  \bra{c} \( \sum_{a,c'} (\log p_{c'} - A) \ketbra{a}{a}_A\otimes\ketbra{c'}{c'} \) \ket{c} 
\\ & =  \underbrace{\tr_A \rho_A^c}_{=1} \( \sum_c p_c \log p_c  - \sum_c p_c A \) 
\\ & = - H(p)  - A . 
\end{align}

The operator appearing in the log in $A_1$ is 
$\sigma \equiv \sum_{c'}p_{c'} \rho_A^{c'} \otimes\ketbra{c'}{c'} $.
Let $ \ket{s^{(c)}_a} $ be eigenvectors of $\rho_A^c$ with eigenvalue $s^{(c)}_i$.  
Then 
$$ \sigma\ket{s^{(c)}_a} \otimes \ket{c} 
= 
\sum_{c'}p_{c'} \rho_A^{c'}  \ket{s^{(c)}_a} \otimes \ket{c'}\underbrace{\vev{c'|c}}_{=\delta_{cc'}}
= p_c s^{(c)}_a  \ket{s^{(c)}_a} \otimes \ket{c} $$
so the eigenvectors of $\sigma$ are 
$$\{ \ket{s^{(c)}_a} \otimes \ket{c} \},~~~~~a=1...\dim\CH_A, c=1...\dim \CH_C. $$
That is,
$$ \sigma = \sum_{a, c}  p_c s^{(c)}_a  \ketbra{s^{(c)}_a}{s^{(c)}_a} \otimes \ketbra{c}{c}. $$
Therefore
\begin{align} A_1 & = - \sum_c p_c  \tr_A \rho_A^c  \bra{c}  \log\sigma \ket{c} 
\\ &= - \sum_c p_c \tr_A \rho_A^c \bra{c} \( \sum_{a,c'} \ketbra{s^{(c')}_a}{s^{(c')}_a}\otimes \ketbra{c'}{c'} \log \( p_{c'}s^{(c')}_a \) \) \ket{c} 
\\ & =  - \sum_c p_c \tr_A \rho_A^c \underbrace{\sum_{a}  \log \( p_c s^{(c)}_a\)   \ketbra{s^{(c)}_a}{s^{(c)}_a} }_{
= \log(p_c) + \log \rho_A^c }
\\ & = - \sum_c p_c \tr_A \rho_A^c \log \rho_A^c - \sum_c p_c \log p_c  \underbrace{\tr\rho_A^c}_{=1} 
\\ & = \sum_c p_c S(\rho_A^c) + H(p). 
\end{align}

Combining the three terms in \eqref{eq:wingedvictorymotherfucker} we have
\begin{align}
S(A|C) & \leq A  - H(p)  - A +  \sum_c p_c S(\rho_A^c) + H(p)
\\ & = \sum_c p_c S(\rho_A^c)  = \overline{S_{X_C}(A)}. 
\end{align}

\section{Conditional mutual information in the Hubbard model}
\label{app:CMI}

The CMI between regions $A$ and $B$ conditioned on $C$ is 
$$I(A:B|C) = S(AC) + S(BC) - S(ABC) - S(C).$$
In this expression, $S(A)$ refers to the entanglement entropy of spins \textit{only} in region $A$, and $S(AB)$ is the total entanglement between spins and charge.  
In general, the CMI is a difference of conditional entropies
\be 
I(A:B|C) = S(AC) - S(C) - (S(ABC)- S(BC))   = S(A|C) - S(AB|C) 
\ee
and in this sense, our discussion here is not independent from 
the analysis of conditional entropy in the main text.
In the case where $ABC$ is the whole system (so $ABC$ is a pure state and $S(AC) = S(B)$ etc), 
the CMI reduces to $I(A:B|C) = I(A:B)$, the ordinary mutual information between $A$ and $B$.

Expectations for the CMI can be found for ergodic and QDL states as in \S\ref{subsec:ce-qdl-vs-ergodic} for various choices of $A,B,C$.  
Of the configurations we have explored, the one which distinguishes them most effectively is the arrangement 
shown in Fig.~\ref{fig:CE-hubbard}, where $A=C, B=D, A<B$.
With this arrangement, in an ergodic state 
\be I(A:B|C) = S(AC) - S(C) - S(AD) + S(D) \sim (A + C) - C - (D+A) +D = 2 A .\ee
In a QDL state, we find 
\be I(A:B|C) = S(AC) - S(C) - S(AD) + S(D) \sim C - C - (D+A) + D = A . \ee
The key effect is the missing $A$ from $S(AC) \sim C$ in the QDL case, which happens
since only the entanglement with $C$ thermalizes $A$.  
This is precisely why conditional entropy distinguishes QDL from ergodic states, as described in \S\ref{subsec:ce-qdl-vs-ergodic}.   

So with this arrangement, both QDL and ergodic states give volume-law behavior of CMI but 
with distinct slopes.
Taking advantage of this effect to compare different states
would require a quantitative understanding of the coefficient of the volume law.
This slope depends on $S(T)$, the entropy at the effective temperature of the state,
which can be extracted from the von Neumann entropy of subsystems.
However, since the effect is in any case not independent of the behavior of the conditional entropy, we 
choose not to pursue this direction further.

\begin{figure}[h!]
 $$
 \includegraphics[width=0.32\textwidth]{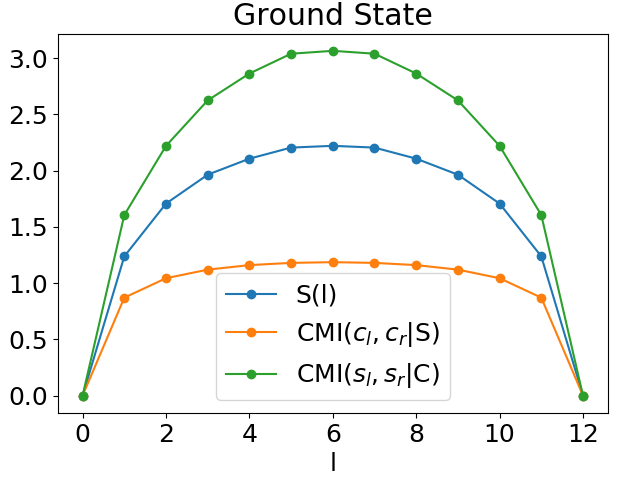}
 \includegraphics[width=0.32\textwidth]{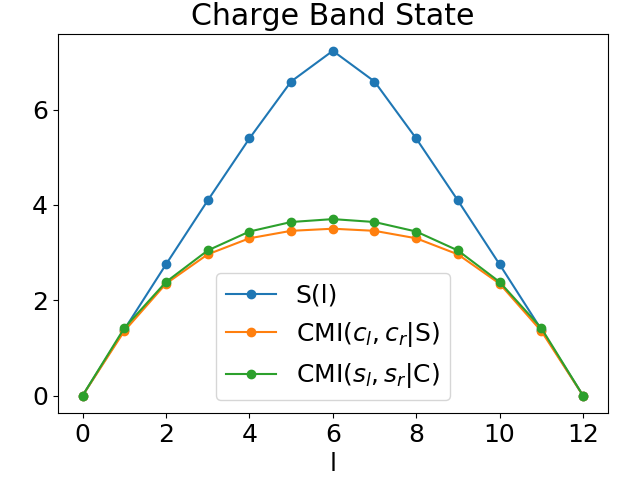}
 \includegraphics[width=0.32\textwidth]{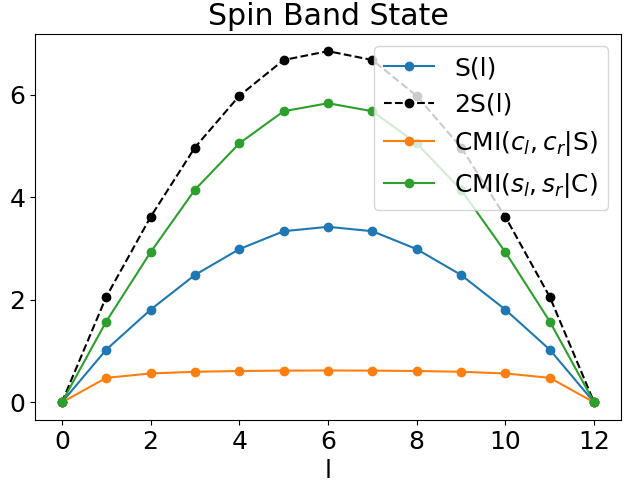}
 $$
 $$\parfig{.3}{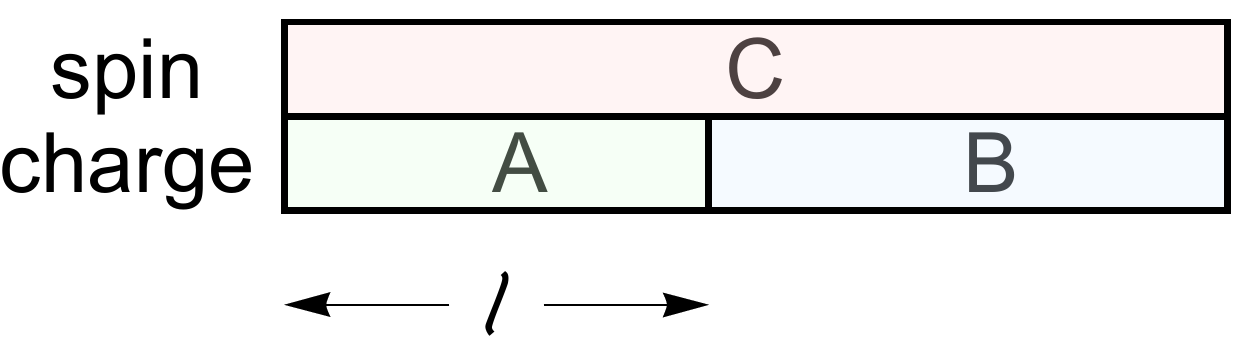} $$
 \caption{Entanglement entropy and CMI measurements $I(A:B|C)$ on a one dimensional Hubbard model as a function of the location of the bipartioning cut at couplings $U = 4,~V = 3/4$. The chain is periodic and has $L=12$ sites, the particle number is at half-filling, and the magnetization is zero.  Left: Ground state. Middle: A generic state in the charge band. Right: A state in the spin band. 
Bottom:  $CMI(c|s)$ refers to CMI with the choice of subsystems indicated.
 $CMI(s|c)$ reverses the roles of spin and charge. 
 }
 \label{fig:hubbard-cmi}
\end{figure}

Results for CMI are presented in Fig.~\ref{fig:hubbard-cmi} for the particular arrangement of $A,B,C$ indicated. 
We calculate CMIs $I(c_l:c_r|s)$, $I(s_l:s_r|c)$ as well as the bipartite von Neumann entropy $S$ for the following three states obtained by exact diagonalization of \ref{eq:hubbard-ham}: the ground state, a generic state taken from the middle of the spectrum, and a state belonging to the `spin band', \ie~a scar state.  

There is indeed a visible difference in the behavior of the CMI between QDL states and ergodic states.
However, a quantitative comparison with general expectations for the behavior of the CMI with the above arrangement of $ABC$ is problematic for the following reason.
Given a partition of a Hilbert space $ABC$ where $\mathcal H_{AB}$ is the same size as $\mathcal H_C$, a finite energy density ergodic eigenstate should have $I(A:B) \sim \sqrt{L_A + L_B}$ after cancellation of the volume law terms \cite{Vidmar:2017pak, Srednicki2019}\footnote{The results of \cite{Vidmar:2017pak, Srednicki2019} follow from an ansatz \cite{Deutsch2010, LuGrover} for the bipartition of an ergodic state
in terms of a wavefunction which is a random matrix, which includes no information about the nature of the bipartition, such as locality.  
(Ref.~\cite{Vidmar:2017pak} matched these results to spatial bipartitions of a non-integrable chain of hardcore bosons.)
The conclusions depend only on the sizes of Hilbert spaces of subsystems with fixed charge.  In particular, \cite{Srednicki2019} generalizes the calculation to include several conserved quantities (the relevant conserved 
quantities here being spin, charge and energy).
As a result, we expect them to apply to our non-local bipartition of an eigenstate of a local Hamiltonian.  
The same conclusions about the leading terms 
would be obtained from a Haar-random pure state \cite{page1993average}.}. 
This represents an area law as a function of $L_A$, while keeping $L_A + L_B$ fixed (as we do in Fig.~\ref{fig:hubbard-cmi}).
The calculation 
of \S\ref{subsec:ce-qdl-vs-ergodic} shows that in a QDL state the CMI for this arrangement is also area law (the extensive terms cancel).
So we attribute the different behavior seen in Fig.~\ref{fig:hubbard-cmi} to non-universal differences in the 
coefficient of the area law.

\bibliographystyle{ucsd}
\bibliography{collection} 
\end{document}